\documentclass[aip, sd, amsmath, amssymb, reprint, groupedaddress]{revtex4-1}
\usepackage{graphicx}
\usepackage{dcolumn}
\usepackage{bm}
\usepackage{xcolor}
\usepackage[normalem]{ulem}
\usepackage[english]{babel}
\usepackage{float}

\newcommand{\com}[1]{{\textcolor{black}{#1}}}

\newcommand{\modea}{\hat{a}^\dagger\hat{a}}
\newcommand{\modeb}{\hat{b}^\dagger\hat{b}}
\newcommand{\coupling}{\hat{a}^\dagger\hat{b} + \hat{a}\hat{b}^\dagger}
\newcommand{\MHz}{$\text{MHz}$}

\newcommand{\avg}[1]{\langle #1 \rangle}
\newcommand{\varg}[0]{{\fontfamily{ptm}\selectfont \textit{g}} }
\newcommand{\ket}[1]{| #1 \rangle}
\newcommand{\bra}[1]{\langle #1 |}

\begin{document}

\preprint{AIP/123-QED}


\title[]{Nearly quantum-limited Josephson-junction Frequency Comb synthesizer}

\makeatletter
\let\@fnsymbol\@fnsymbol@latex
\@booleanfalse\altaffilletter@sw
\makeatother

\author{Pinlei Lu*}
\affiliation{Department of Physics and Astronomy, University of Pittsburgh}
\author{Tzu-Chiao Chien}
\affiliation{Department of Physics and Astronomy, University of Pittsburgh}
\author{Xi Cao}
\affiliation{Department of Physics and Astronomy, University of Pittsburgh}
\author{Olivia Lanes}
\affiliation{Department of Physics and Astronomy, University of Pittsburgh}
\author{Chao Zhou}
\affiliation{Department of Physics and Astronomy, University of Pittsburgh}
\author{Saeed Khan*, Hakan E. T\"ureci}
\affiliation{Department of Electrical Engineering, Princeton University \\ *These authors contributed equally to this publication.}
\author{Michael J. Hatridge}
\affiliation{Department of Physics and Astronomy, University of Pittsburgh}


\date{\today}

\begin{abstract}
While coherently-driven Kerr microcavities have rapidly matured as a platform for frequency comb formation, such microresonators generally possess weak Kerr coefficients; consequently, triggering comb generation requires millions of photons to be circulating inside the cavity.  This suppresses the role of quantum fluctuations in the comb's dynamics. In this paper, we realize a minimal version of coherently-driven Kerr-mediated microwave frequency combs in the circuit QED architecture, where the quantum vacuum's fluctuations are the primary limitation on comb coherence. We achieve a comb phase coherence of up to 35~$\mu$s, approaching the theoretical device quantum limit of 55~$\mu$s, and vastly longer than the modes' inherent lifetimes of 13~ns. The ability within cQED to engineer stronger nonlinearities than optical microresonators, together with operation at cryogenic temperatures, and excellent agreement of comb dynamics with quantum theory indicates a promising platform for the study of \com{complex dynamics of quantum nonlinear systems}.
\end{abstract}

\maketitle


\com{\section*{Introduction}}

While the circuit QED (cQED) architecture has built its success on strongly-coupled qubit-cavity experiments~\cite{Blais2004, Girvin2009, GirvinNote, martinis_quantum_2020, Blais2020}, it has also been firmly established as a versatile platform to realize a broader variety of quantum nonlinear systems~\cite{Haroche2020}.  Josephson-junction junction based superconducting circuits have also enabled devices from quantum-limited amplifiers~\cite{Vijay2009, yamamoto_2008_fluxJPA, castellanos-beltran_2007, aumentado_2020_amplifiers, roy_introduction_2016} and single-microwave photon detectors~\cite{Walsh2017, inomata_single_2016, Narala_2016, poudel_2012} with application ranging from quantum information processing to the search for dark matter axions, to hybrid quantum systems~\cite{Clerk2020}. A key factor determining the breadth of realizable quantum nonlinear devices, and thus feasibility of future applications, is understanding the diverse dynamical regimes enabled by Josephson-junctions.

A nonlinear dynamical regime that has yet to be realized via a Josephson-junction mediated Kerr nonlinearity is that of frequency comb formation. Distinct from Kerr-nonlinear amplifiers which operate in regimes with at least one classically stable fixed point in phase space, frequency comb formation is marked by a system undergoing stable periodic excursions around \textit{unstable} fixed points. In the optical domain, coherently-driven microresonators utilizing the Kerr nonlinearity have emerged as the leading platform for frequency comb generation~\cite{Haye2007, Haye2008, Levy2010, herr_universal_2012, Herr2013, kippenberg_dissipative_2018}; however the typically weak Kerr nonlinearity of optical microresonators~\cite{gaeta_photonic-chip-based_2019} means that contemporary comb generation requires $\sim\mu$W power input\cite{stern_battery-operated_2018}, corresponding to millions of circulating cavity photons~\cite{kues_quantum_2019}.  Similar results have been achieved in superconducting circuits using the weak nonlinearity of kinetic-inductance in very long resonators\cite{pappas_frequency_2014}.  As a result, vacuum fluctuations amplified by the comb-generating nonlinear process are much weaker in comparison~\cite{newbury_noise_2007}.

\begin{figure*}[t]
	\includegraphics[scale = 1.0]{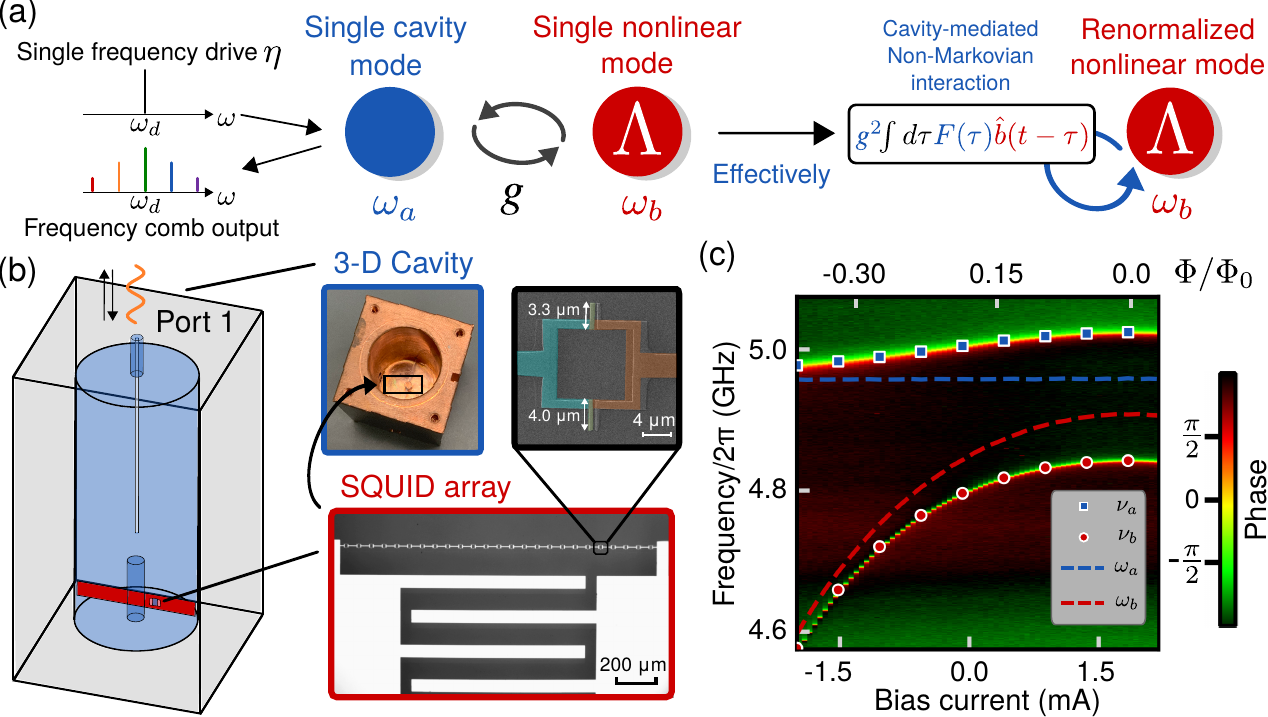}
	\caption{\textbf{System schematic and flux sweep.} (a) The two-mode device can be represented by a single cavity mode (blue, $\omega_a$) coupled linearly to a single Kerr mode (red, $\omega_b$).  We create a comb in this device by driving with a carefully chosen, single microwave drive of strength $\eta$ at frequency $\omega_d$, which interacts with the device to create a series of tones at regularly spaced output frequencies (the `comb'). The linear mode plays the role of mediating a delayed self-interaction of the nonlinear mode (right panel), with kernel $F(\tau) = e^{-\tau/\chi_a}$. In the strong coupling regime, the interaction's non-Markovian nature fundamentally modifies the nonlinear mode's stability, enabling comb formation. (b) Circuit QED implementation of the schematic in (a). The nonlinear mode inductance is formed by 25 Superconducting QUantum Interference Devices (SQUIDs) in series; False-color SEM image of a component SQUID indicates the small asymmetry employed to alleviate hysteresis. The SQUID array is coupled to an antenna that both forms the capacitance of the nonlinear mode and controls its dipole coupling with the linear mode; the latter is the $\lambda/4$ mode of a coaxial 3-D cavity, fabricated of copper to permit the passage of an external DC flux that threads all the SQUIDs. The input signal drives the cavity through port 1, and $S_{11}(\omega)$ is monitored. (c) Color plot of reflected signal vs. frequency from port 1 ($S_{11}(\omega)$) for a range of applied coil bias currents/applied SQUID fluxes.  The linear and nonlinear mode frequencies are highlighted by blue squares and red dots respectively. By fitting for the bare mode frequencies, we determine $\varg/2\pi=87.6956$ MHz, as well as the bare mode frequencies, represented by the dashed lines.}
	\label{fig:schematic}
\end{figure*}

In this paper, we harness the Josephson junction to realize a minimal version of Kerr-mediated microwave frequency combs based on a recent theoretical proposal~\cite{Saeed2018}. Our minimal realization within cQED consists of just two coupled modes, of which only one possesses a Kerr nonlinearity furnished by Josephson junctions, as shown in Fig.~\ref{fig:schematic}. Although our device is based on familiar cQED components, it operates in a distinct regime within the landscape of nonlinear cQED devices: while strongly-coupled like transmon-cavity systems~\cite{Koch2007}, its nonlinearity is in fact weaker and is operated under much stronger driving. On the other hand, the device exhibits stronger couplings yet smaller detunings and weaker drives than Kerr-mediated bifurcation and parametric amplifiers~\cite{siddiqi_rf-driven_2004, Vijay2009}. This allows us to realize a \sout{novel} unstable regime where a single frequency drive tone generates coherent frequency combs over a large parameter space.

Crucially, the strong engineerable nonlinearities in cQED and operation at cryogenic temperatures brings quantum fluctuations to the fore ahead of thermal and dephasing effects in our comb synthesizer: the phase coherence of the generated combs is fundamentally limited by vacuum fluctuations that are amplified by the nonlinear comb-generating process itself. A microscopic nonlinear quantum theory of our two-mode device, in addition to providing precise operating parameters for this comb-generating regime, enables us to quantify this quantum limit on comb phase coherence. By also characterizing and explaining the dependence of coherence on operating parameters like detuning and drive power, we provide a detailed quantitative study of the phase coherence of frequency combs near the quantum limit.

Built up of fundamentally quantum components, we believe this highly-controllable cQED realization can serve as a necessary building block for Kerr-nonlinear systems operating in classically-unstable \textit{and} deeply-quantum regimes, exploring dynamics beyond coherent frequency comb formation. For certain parameter regimes, our device exhibits temporal instabilities marked by large, irregular excursions in phase space, distinct from regular comb dynamics and reminiscent of chaos. More excitingly, while our work indicates that strong quantum fluctuations limit the coherence of generated frequency combs, they are also features of deeply-quantum regimes necessary for displaying quantum effects such as squeezing, entanglement, and generation of non-Gaussian states. Our realization thus marks a promising first step in exploring the potentially competing role of strong quantum fluctuations in quantum dynamics within classically unstable regimes.

\section*{Theory and results}

\textbf{System schematic and device overview.} The Hamiltonian of our device consists of a linear mode $\hat{a}$ with uncoupled resonant frequency $\omega_a$, linearly coupled with strength $\varg$ to a nonlinear mode $\hat{b}$ with uncoupled resonant frequency $\omega_b$; see Fig.~\ref{fig:schematic}(a). The linear mode is driven by a coherent tone with frequency $\omega_d$ and amplitude $\eta$, and the system Hamiltonian in the frame rotating with this drive takes the form:
    \begin{equation}
        \begin{split}
           \hat{\mathcal{H}}/\hbar = &\  -\Delta_{da} \modea -\Delta_{db} \modeb  - \frac{\Lambda}{2} \hat{b}^{\dagger}\hat{b}^{\dagger}\hat{b}\hat{b} \\
            &  + \varg (\coupling) + \eta (\hat{a} + \hat{a}^\dagger)     
        \end{split}
        \label{eq:hsys}
    \end{equation}
where $\Delta_{da/db} = \omega_d - \omega_{a/b}$ and $\Lambda > 0$ is the strength of the Kerr nonlinearity. In our experiment (Fig.~\ref{fig:schematic}(c)), the nonlinear mode is realized as a Superconducting QUantum Interference Device (SQUID)\cite{squidHandBook} array (Device A: 25 SQUIDs; Device B: 5 SQUIDs).  The SQUIDs act together as a flux-tunable, nonlinear inductor, which is shunted with a planar interdigitated capacitor/antenna to form a nonlinear microwave mode.   Weakly asymmetric SQUIDs (with critical current ratio of 1.2:1) are used to build up the array, alleviating otherwise large hysteresis effects at the cost of a reduction in tunability of the nonlinear mode frequency\cite{hutchings_tunable_2017}. The device is deposited on a sapphire substrate and capacitively coupled to the $\lambda/4$ mode of a coaxial 3-D copper cavity~\cite{paik_transmon_2011}. This driven-dissipative system is then described by the master equation: $\dot{\hat{\rho}} = -i[\hat{\mathcal{H}},\hat{\rho}] + \kappa\mathcal{D}[\hat{a}]\hat{\rho} + \gamma\mathcal{D}[\hat{b}]\hat{\rho} + \gamma_{\varphi}\mathcal{D}[\hat{b}^{\dagger}\hat{b}]\hat{\rho}$, which includes linear damping rates $\kappa$ ($\gamma$) for modes $\hat{a}$ ($\hat{b}$), and pure dephasing ($\gamma_{\varphi}$) for the flux-tunable nonlinear mode; thermal fluctuations are neglected. By sweeping the flux through the SQUIDs to tune the nonlinear mode frequency, and making a measurement of the reflection coefficient $S_{11}(\omega)$, we extract (Fig.~\ref{fig:schematic}(c)) a coupling strength of $\varg/2\pi = 87.6956$ \MHz~between the modes, and linear mode damping rate $\kappa/2\pi = 10.9308$ \MHz. Via pump-probe measurements~\cite{SI} we also extract a Kerr nonlinearity of $\Lambda/2\pi = 5.96~$kHz, such that $\Lambda/\kappa \sim 10^{-3}$, stronger than typical values of $\sim 10^{-5}$ for optical microresonators~\cite{gaeta_photonic-chip-based_2019, SI}.

\begin{table}[b]
\begin{tabular}{|l|l|l|}
\hline
    & Device A (25 SQUIDs) & Device B (5 SQUIDs) \\ \hline
$\omega_{b}/2\pi (\text{GHz})$ & 4.956806 & 4.951073 \\ \hline
$\varg/2\pi (\text{MHz})$ & 87.6956 & 89.25 \\ \hline
$\Lambda/2\pi (\text{MHz})$ & $5.96 \times 10^{-3}$ & $152.6\times 10^{-3}$ \\ \hline
$\kappa/2\pi (\text{MHz})$  & 10.9308 & 22.84 \\ \hline
\end{tabular}
\caption{\textbf{Device parameters.} Coupling strength $\varg$, nonlinearity $\Lambda$, and bare cavity damping rate $\kappa$ for Device A (25 SQUIDs) and Device B (5 SQUIDs); nonlinearity suppression by a factor $\sim 25$ is measured, as designed.}
\end{table}

\textbf{Comb generation and phase diagram.} Analysis of this system in Ref.~\cite{Saeed2018} showed that the linear mode effectively equips the nonlinear mode with a delayed self-interaction (see Fig.~\ref{fig:schematic}(a)), whose influence is dictated by the coupling $\varg$ and the linear mode susceptibility $\chi_a = (-i\Delta_{da}+\frac{\kappa}{2})^{-1}$. Under suitable coupling, drive, and detuning conditions, this two-mode system can go beyond typical bifurcation dynamics associated with Kerr nonlinear devices to exhibit frequency comb formation.  To illustrate this, we plot the classical phase diagram for measured Device A parameters in  Fig.~\ref{fig:varyingWQ}(a), as a function of drive detunings $\Delta_{da}, \Delta_{db}$~\com{(see Appendix~\ref{app:classical})}. For each pair of detunings, we consider a range of experimentally accessible drive powers (-132 dBm to -67 dBm), and classify phases according to the number of fixed points (FPs) and stable fixed points (SFPs) observable within this driving range. For large $|\Delta_{da}|$ (small $|\chi_a|$) relative to $\varg$, only two types of phases are exhibited: blank regions, where the system admits one SFP for all driving powers considered, or hatched regions, where for some subset of driving powers, three FPs exist. In either case, at least one fixed point is always stable~\cite{Saeed2018}. These phases are reminiscent of the standard Kerr bistability, and unsurprisingly so: in this regime, the effective coupling $\varg|\chi_a|$ is weak, and the mediated interaction may be treated within a Markov approximation.

However, for intermediate $|\Delta_{da}|$ such that $\varg|\chi_a| \gtrsim 1$ (on resonance, we require $\varg > \kappa/2$, comfortably satisfied by Device A), the non-Markovian nature of the interaction manifests in a qualitative change of the nonlinear mode's stability, marked by regions (shaded red) where no stable fixed points exist for a subset of the driving powers considered. Here, classical Lyapunov analysis reveals the possibility of our device exhibiting stable limit cycles with period $T = \frac{2\pi}{\Delta}$ and comb-like frequency spectra with spacing $\Delta$, and even chaotic dynamics deeper into the unstable regime\com{, namely at more negative detunings and stronger drive powers~(see Appendix~\ref{app:lyapunov})}.

\begin{figure}[t]
	\includegraphics[scale = 1.0]{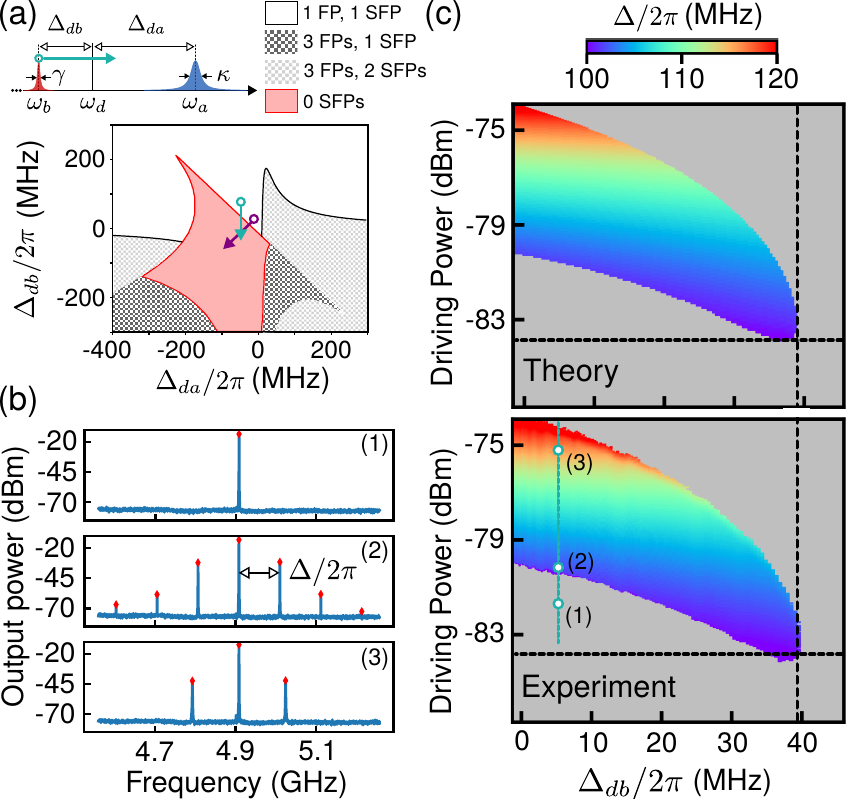}
	\caption{\textbf{Phase diagram and comb spectrum}. (a) Theoretically predicted phase diagram in $\Delta_{db}$-$\Delta_{da}$ space, indicating observable phases characterized by number and stability of classical fixed points (FPs) over a range of experimentally accessible drive powers (-132 dBm to -67 dBm, see text for details). Here, the unstable regime (shaded red) exhibiting no stable fixed points (SFPs) is entered by fixing the drive frequency to $\omega_d = 4.9085(2\pi)$ GHz, varying the nonlinear mode frequency along the direction of the green arrow via a flux sweep (see schematic), and observing the output power spectra. (b) Typical power spectra as a function of increasing drive power, along the indicated cross section of the experimental phase diagram in (c). At low powers, the system exhibits single frequency output at $\omega_d$ (1), marked gray in the phase diagram. With increasing drive power (2), the system enters a regime with regularly spaced multifrequency output, with spacing $\Delta$ which is plotted in the phase diagram in Drive-$\Delta_{db}$ space. With stronger driving power (3), the spacing $\Delta$ increases, while fewer side peaks are observed above the noise background, before the system ultimately exits the unstable regime and single frequency output resumes. The theoretical phase diagram is plotted in the top panel of (c) for comparison.}
	\label{fig:varyingWQ}
\end{figure}

To observe the response of our quantum device in this rich dynamical regime, we enter the unstable region along the green arrow in Fig.~\ref{fig:varyingWQ}~(a), by fixing the drive frequency so that $\Delta_{da}/2\pi = -47.8$ \MHz, and flux tuning the nonlinear mode frequency. In search of the frequency domain signature of comb formation, we measure the frequency response in drive-$\Delta_{db}$ parameter space using a spectrum analyzer, with typical results at fixed $\Delta_{db}$ shown in Fig.~\ref{fig:varyingWQ}(b). At low powers (1), the system exhibits a single frequency response at the drive frequency, corresponding to the stable fixed point. However, as the power is increased, a multifrequency spectrum emerges with equidistant peaks (2 and 3). The spacings $\Delta$ extracted from these power spectra are used to construct the experimental phase diagram in Fig.~\ref{fig:varyingWQ}(c), with the theoretical result over the same parameter space provided for comparison. We find remarkable agreement between theory and experiment; only a single fitting offset is used to account for scaling factors along the drive power axis. 

\textbf{Temporal coherence and dynamical response.} Power spectrum measurements provide a key signature of comb formation but are insensitive to the nontrivial phase dynamics of these complex nonlinear solutions. While the central comb peak has a definite phase set by the incident coherent tone, the relative phase $\theta(t)$ of generated sidebands relative to the central peak phase is free to diffuse~\cite{ablowitz_noise-induced_2006, navarrete-benlloch_general_2017}. This diffusion sets the comb linewidth and thus provides the ultimate limit to any precision measurements made using the comb in question~\cite{coluccelli_frequency-noise_2015}. To quantify the phase coherence, we measure the \textit{steady-state} first-order temporal coherence function $G^{(1)}(\tau)$, defined as~\cite{da_silva_schemes_2010}:

\begin{align}
    G^{(1)}(\tau) = \lim_{t\to\infty}\frac{\avg{I(t)I(t+\tau)} - \avg{I(t)}^2}{\avg{I(t)^2}-\avg{I(t)}^2}
    \label{eq:g1}
\end{align}

\begin{figure*}[t]
	\includegraphics[scale = 1.0]{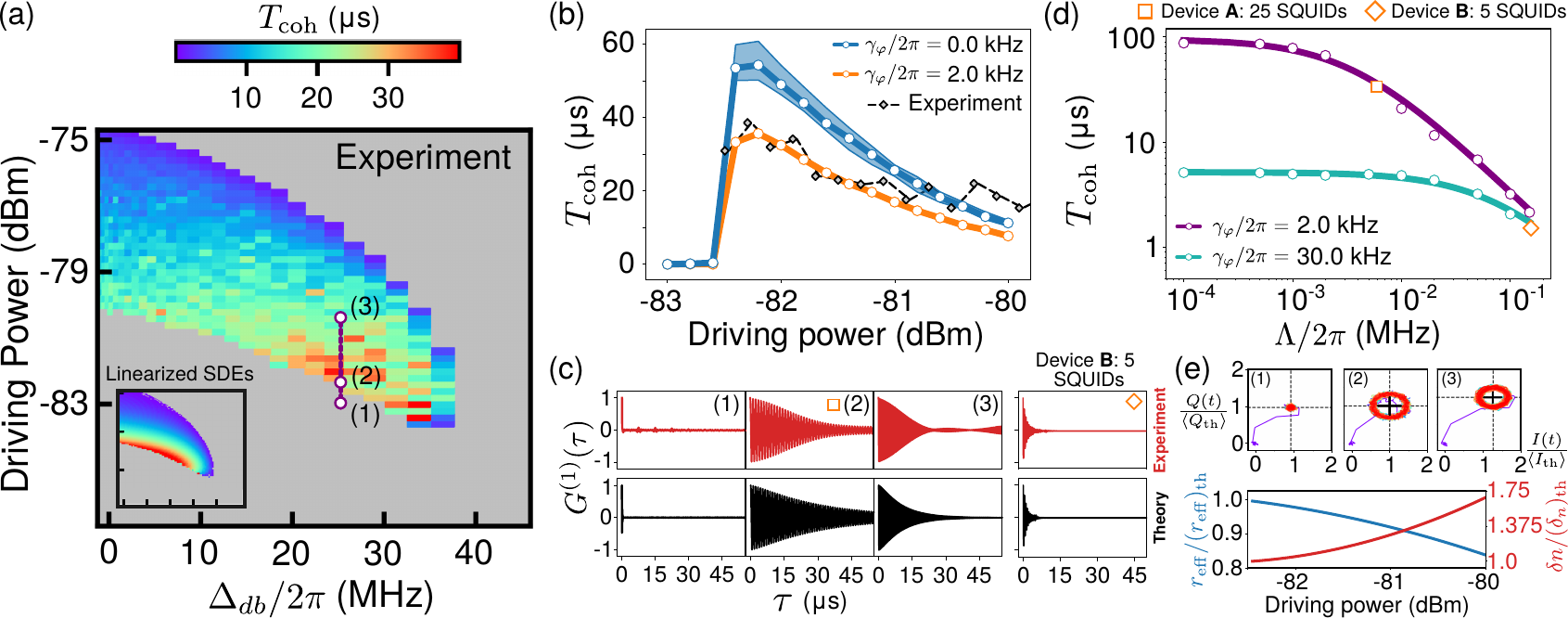}
	\caption{\textbf{Comb coherence.} (a) Coherence time $T_{\rm coh}$ extracted by measuring $G^{(1)}(\tau)$ (Eq.~\ref{eq:g1}) for the same operating parameters as Fig.~\ref{fig:varyingWQ} b. Inset: approximate $T_{\rm coh}$ calculated using Floquet analysis of linearized SDEs (see text). (b) Cross section of phase diagram along the dashed line in a, in black. The blue curve and shaded region indicates the theoretically calculated coherence time due to nonlinearity alone (pure dephasing $\gamma_{\varphi}=0$). The orange curve shows $T_{\rm coh}$ for $\gamma_{\phi} = 2.0(2\pi)~$kHz, showing good agreement with the experimental result. (c) Experimental (top panel) and theoretical (lower panel) $G^{(1)}(\tau)$ for $\gamma_{\varphi} = 2.0(2\pi)~$kHz at positions 1 (stable regime), 2 (threshold of comb formation), and 3 (higher drive power in comb regime). (d) Numerical results for variation of coherence time with nonlinearity. Purple and green points correspond to parameters for Device A (25 SQUIDs) and B (5 SQUIDs) respectively, obtained by varying the nonlinearity alone. Experimental results for the devices are shown by the orange square and diamond; corresponding $G^{(1)}(\tau)$ are marked by the same symbols in (c). Curves are fits to $T_{\rm coh} = a(\gamma_{\varphi} + b\Lambda)^{-1}$, with $(a,b) = (1.19, 0.55)$ for Device A and $(0.94, 0.40)$ for Device B. (e) Top panel: $I$-$Q$ trace at positions \{1,2,3\}, showing 2-D projection of the limit cycle orbit which decreases in radius with increasing power. Lower panel: theoretical effective radius of limit cycle $r_{\rm eff}$ (solid blue) and standard deviation of noise projected tangential to limit cycle, $\delta n$ (solid red, right hand axis)\com{, scaled by their values at threshold}. The \com{relative} decrease in $r_{\rm eff}$ combined with the increase in $\delta n$ point towards a reduction in coherence time with increasing drive power (see text).}
	\label{fig:coherence}
\end{figure*}

To do so, we first obtain the time-domain cavity output $I(t)$ using a single side band (SSB) mixer to downconvert the dominant sideband peak to around the 100 MHz regime, followed by homodyne detection via a 500 MSample/s digitizer to demodulate the output signal, and finally compute its time-domain autocorrelation. The normalized coherence function $G^{(1)}(\tau)$ decays from its maximum value of unity (at $\tau = 0$) towards $G^{(1)}(\tau) = 0$ over a time scale $T_{\rm coh}$ determined by the loss mechanisms affecting the system dynamics. We measure $G^{(1)}(\tau)$ in the parameter space explored in Fig.~\ref{fig:varyingWQ}(c), and extract $T_{\rm coh}$ as the decay constant of the observed function envelopes; the results are plotted in Fig.~\ref{fig:coherence}(a). Focusing in particular on the indicated cross-section at $\Delta_{db}/2\pi = 25.2~$MHz, we plot the measured $G^{(1)}(\tau)$ functions at positions $\{1,2,3\}$ in the top panel of Fig.~\ref{fig:coherence}(c). Outside the comb regime (1), $G^{(1)}(\tau)$ decays on a timescale of $\sim 13~$ns, set by the fastest decay rate, namely the bare cavity loss $\kappa$. However, a qualitative change is observed in $G^{(1)}(\tau)$ when the system transitions into the comb regime (2), with a sharp increase in coherence time to a maximum of $36.7~\mu$s, significantly longer than the timescale set by $\kappa$. This observation, together with the decrease in $T_{\rm coh}$ with increasing drive power (3), highlights a key feature of the self-oscillating regime: the intrinsic energy loss of the system is overcome and coherence is therefore no longer determined by the bare energy loss rates.

This naturally raises the question: what limits the observed phase coherence? The answer lies in the full quantum description of the strongly-driven, weakly nonlinear two-mode system. In this regime, we employ a phase-space approach based on the Positive-$P$ representation~\cite{Saeed2018, carmichael_statistical_2002, drummond_generalised_1980}, obtaining a set of stochastic differential equations (SDEs) for phase space variables $\vec{\zeta} = (\alpha,\alpha^{\dagger},\beta,\beta^{\dagger})^T$ associated with operators $(\hat{a},\hat{a}^{\dagger},\hat{b},\hat{b}^{\dagger})^T$. The SDEs take the general form \com{(see Appendix~\ref{app:prep})}:
\begin{align}
    d\vec{\zeta}(t) = \vec{A}_{\rm c}(\vec{\zeta})~dt + \mathbf{B}_{\rm st}(\vec{\zeta},\Lambda,\gamma_{\varphi}) d\vec{W}(t) 
    \label{eq:sdes}
\end{align}
The deterministic contribution ($\propto \vec{A}_{\rm c}$) describes noise-free classical dynamics of the two-mode system, which yields perfectly coherent combs. The remaining stochastic terms $\propto d\vec{W}(t)$ (vector of independent Wiener increments) then describe deviations from classical dynamics, here including fluctuations due to the quantum nonlinearity $\Lambda$ and pure dephasing $\gamma_{\varphi}$. These fluctuations are ultimately responsible for phase diffusion that limits comb coherence. The stochastic terms take the explicit form $\mathbf{B}_{\rm st}(\vec{\zeta},\Lambda,\gamma_{\varphi})d\vec{W}(t) = \sqrt{\Gamma} \mathbf{B}_1(\vec{\zeta}) d\vec{W}_1(t) + \sqrt{\gamma_{\varphi}} \mathbf{B}_2(\vec{\zeta}) d\vec{W}_2(t)$, where $\Gamma = \sqrt{\Lambda^2+\gamma_{\varphi}^2}$. Crucially, we note that even in the absence of pure dephasing, $\gamma_{\varphi} \to 0$, the stochastic terms do not vanish: a contribution due to the intrinsic nonlinearity of the system always remains, setting a fundamental limit on comb coherence. This is verified by simulating Eqs.~(\ref{eq:sdes}) for $\gamma_{\varphi} = 0$ and the experimentally measured nonlinearity of $\Lambda/2\pi = 5.96~$kHz, and obtaining $T_{\rm coh}$\cite{maillet_classical_2016, aspelmeyer_cavity_2014}; the results are shown by the blue curve in Fig.~\ref{fig:coherence}(b), with the blue shaded region being a 95\% confidence bound accounting for uncertainty in $\Lambda$. The maximum $T_{\rm coh}$ is thus limited to around 55~$\mu$s by amplified quantum fluctuations due to the device nonlinearity alone under these operating conditions. This of course exceeds the maximum observed $T_{\rm coh}$ since $\gamma_{\varphi} \neq 0$. For $\gamma_{\varphi}/2\pi \simeq 2.0~$kHz (orange) we find good agreement with experiment (gray) \com{(see Appendix~\ref{app:nlDephasing})}; simulated $G^{(1)}(\tau)$ at positions $\{1,2,3\}$ are shown (Fig.~\ref{fig:coherence}~c, black) for comparison. The relatively small $\gamma_{\varphi}$ is not unexpected given both the narrow modulation range of the asymmetric SQUID array~\cite{hutchings_tunable_2017} and operation at $\Phi/\Phi_0 \lesssim 0.12$, close to the flux noise sweet spot (see Fig.~\ref{fig:schematic}~\com{(c)}).

Since $\Lambda$ cannot be varied $\textit{in-situ}$ while holding other parameters fixed, we confirm its influence on $T_{\rm coh}$ by employing Device B; this 5-SQUID device is engineered to have the same total inductance as Device A, while possessing a 25-fold stronger nonlinearity~\cite{Eichler2014} of $\Lambda/2\pi = 152.6~$kHz. While we obtain similar multifrequency behaviour (full results in SI~\cite{SI}), coherence times for this device are much shorter, $T_{\rm coh} \lesssim 1.5~\mu$s (see Fig.~\ref{fig:coherence}(c) for measured and simulated $G^{(1)}(\tau)$ at typical operating parameters). Although Device B is operated away from the flux-noise sweet spot~\cite{SI}, and thus experiences a larger estimated $\gamma_{\varphi}/2\pi \simeq 30.0~$kHz, we find that its much stronger nonlinearity is dominant in limiting comb coherence. To confirm the dependence of $T_{\rm coh}$ on $\Lambda$ and $\gamma_{\varphi}$ numerically, we simulate $T_{\rm coh}$ at fixed positions on the phase diagrams of both devices, while varying $\Lambda$. The results are plotted in Fig.~\ref{fig:coherence}(d), in purple (green) for Device A (Device B) parameters, with the experimental result indicated by the square (diamond). They are well described by fits to $T_{\rm coh} = a(\gamma_{\varphi} + b\Lambda)^{-1}$ (curves); we find $b =({\rm A\!:\ }0.40,{\rm B\!:\ }0.55) \neq 1$, consistent with $\Lambda$ and $\gamma_{\varphi}$-contributions to dephasing originating from different stochastic terms in Eqs.~(\ref{eq:sdes}). More importantly, both devices clearly operate in the regime where $b \Lambda \gtrsim \gamma_{\varphi}$, and thus $T_{\rm coh}$ is predominantly set by the nonlinearity.
    
\begin{figure}[t]
	\includegraphics[scale = 1.0]{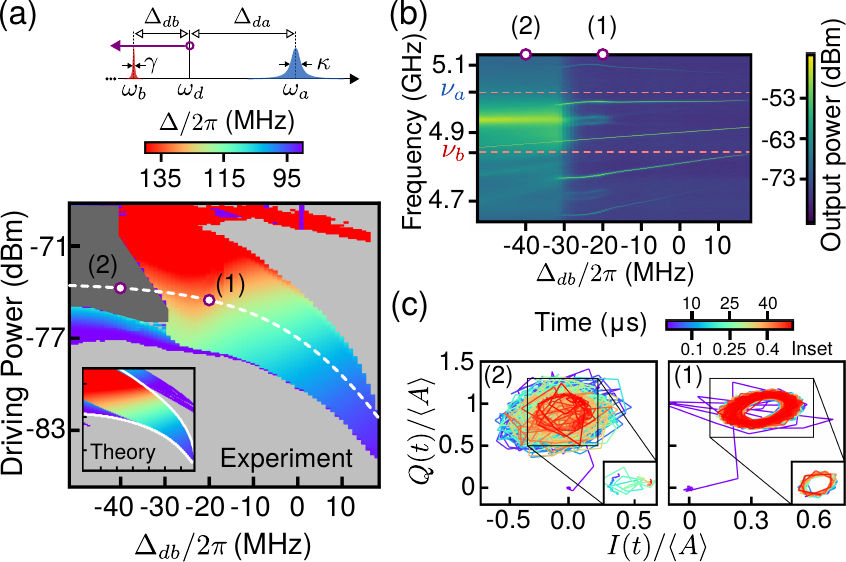}
	\caption{\textbf{Temporal instabilities.} (a) By fixing $\omega_a, \omega_b$ and varying $\omega_d$ (top panel), the system can be driven to the regime with no stable fixed points along the purple arrow in Fig.~\ref{fig:varyingWQ}~(a), while leaving the underlying mode structure unchanged. The resulting phase diagram plotting observed comb spacing $\Delta$ is shown, with the theoretical prediction in the inset. For $\Delta_{db}/2\pi \lesssim -30~$MHz and strong enough driving, a distinct regime emerges (dark gray) where the output spectrum broadens significantly. In (b), we show the typical evolution of the spectrum across the white dashed line, chosen to show a large variation in comb spacing. In dashed red are the underlying polariton resonances, indicating that the emergent comb peaks do not exactly coincide with these resonances. (c) Time dynamics as observed via $I$-$Q$ traces, with both axes scaled by $\avg{A} = \sqrt{\avg{I^2} + \avg{Q^2}}$ at position 1, for ease of direct comparison. In the stable comb regime (1), the cavity response settles into an obvious orbit as before; the inset shows a 500~ns trace after $t = 40~\mu$s, demonstrating the stable orbit. In the unstable regime (2), the response shows large deviations over time and no periodic phase space trajectory is observed.}
	\label{fig:varyingWD}
\end{figure}

However, as observed in Fig.~\ref{fig:coherence}(a), $T_{\rm coh}$ also depends nontrivially on \textit{operating} parameters (e.g. drive power, detuning), even if $\Lambda$, $\gamma_{\varphi}$ are held fixed. This dependence is intimately related to the nature of the dynamical comb regime, where the system traverses a periodic trajectory in phase space. The shape of this trajectory, which changes with operating parameters, controls its susceptibility to noise, as well as the noise itself when the latter is \textit{multiplicative} (dependent on $\vec{\zeta}(t)$, as $\mathbf{B}_{\rm st}$ is). This connection can be made precise via a linearized Floquet analysis~\cite{demir_phase_2000, navarrete-benlloch_general_2017, drummond_quantum_2003} of the SDEs around the \textit{classical} limit cycle trajectory $\vec{\zeta}_{\rm c}(t)$. In this weak-fluctuations approach~\cite{gardiner_input_1985, SI}, the phase $\theta(t)$ of the limit cycle solution evolves according to the SDE: $r_{\rm eff} \dot{\theta} = n(t)$, and the coherence time $T_{\rm coh}$ can be related to the variance of this diffusing phase, \com{$T_{\rm coh}^{-1} \propto \avg{[\theta(T)-\theta(0)]^2}$~(see Appendix~\ref{app:floquet})}. Here $r_{\rm eff}$ is the effective limit cycle radius, defined via $r_{\rm eff}\Delta = \sqrt{\frac{1}{T}\int_0^T dt~||\vec{v}(t)||^2}$ where $\vec{v}(t) = \dot{\vec{\zeta}}_{\rm c}(t)$ is the tangential velocity of limit cycle traversal. Secondly, $n(t)$ is the projection of stochastic terms $\mathbf{B}_{\rm st}(\vec{\zeta}_{\rm c}(t)) d\vec{W}$ onto the limit cycle trajectory. Noise projected onto the limit cycle therefore provides an impulse that causes $\theta(t)$ to diffuse, while $r_{\rm eff}$ provides an inertial term: the larger the radius, the more $\theta(t)$ resists diffusion. We plot the average projected noise standard deviation, $\delta n = \sqrt{\frac{1}{T}\int_0^T dt~\avg{n(t)^2}}$ and the effective limit cycle radius $r_{\rm eff}$ along the indicated cross-section of Fig.~\ref{fig:coherence}(a), scaled by their values at the threshold of comb formation. The limit cycle radius (blue) decreases with increasing power; this is also seen experimentally in $I$-$Q$ traces (top panel), positions 2 to 3, which can be viewed as a 2-D Poincar\'e section of the limit cycle trajectory. Additionally, the noise strength $\delta n$ (red, right hand axis) increases, in a clear manifestation of its multiplicative nature. Both effects tend to reduce $T_{\rm coh}$, as captured by both the linearized analysis (Fig.~\ref{fig:coherence}(a), inset) and full SDE simulations (Fig.~\ref{fig:coherence}(b)).

\com{Finally, we note that multiplicative noise can also manifest in non-exponential decay of phase coherence. However, for the operating parameters explored in Fig.~\ref{fig:coherence}, theoretical simulations predict deviations from exponential decay to be minimal~\cite{SI}, and experimentally observed weak non-exponential signatures (such as (3) in Fig.~\ref{fig:coherence}(c)) can be attributed to electronic $1/f$ noise~\cite{SI}. With the use of additional probing systems and judicious choice of operating parameters\cite{Bylander2011, Yan2013, Andersson2019}, this system could be used to study non-exponential phase decoherence due to quantum fluctuations.}

\textbf{Temporal instabilities and further explorations.} While we have demonstrated the formation of stable frequency combs with this minimal two-mode Kerr system, even more complex dynamical phenomena may be observed deeper in the regime with no stable fixed points. We explore this region by fixing $\omega_b = 4.91~{\rm GHz}$ and varying $\omega_d$ instead, now entering the unstable region along the purple arrow in Fig.~\ref{fig:varyingWQ}(a). The experimental phase diagram in Fig.~\ref{fig:varyingWD}(a) plots spacings $\Delta$ where combs are observed, together with a dark gray region where the spectrum no longer exhibits a comb. The typical variation in spectrum is shown in Fig.~\ref{fig:varyingWD}(b). For $\Delta_{db}/2\pi \gtrsim -30~$MHz, a clear comb spectrum is observed with a spacing that varies with $\omega_d$; the system polariton frequencies $\nu_a, \nu_b$ (unchanged with $\omega_d$) are marked in dashed pink, confirming that comb peaks do not always coincide with passive modes of the two-mode system. 

For $\Delta_{db}/2\pi \lesssim -30~$MHz, the spectrum abruptly changes, exhibiting a single broad peak and an increased noise background. Analyzing $I$-$Q$ traces in Fig.~\ref{fig:varyingWD}(c), dynamics in this region (2) show large deviations with time and while recurringly confined to a region of phase space do not follow a regular trajectory, even on short timescales (inset), in stark contrast to regular periodic dynamics for stable comb operation (1). Note that these temporal instabilities disagree with results of a weak quantum fluctuations analysis in this regime (Fig.~\ref{fig:varyingWD}(a), inset), which simply predicts frequency combs with finite coherence akin to Fig.~\ref{fig:coherence} (although instabilities do manifest for more negative detunings~\cite{SI}). Curiously, quantum dynamics here are also too complex to be captured by simulating the exact SDEs in Eqs.~(\ref{eq:sdes}), which run into familiar numerical difficulties encountered in the application of phase-space stochastic approaches to strong-quantum systems~\cite{gilchrist_prep}. This could be indicative of qualitative deviations from classically stable limit cycles not captured by a linearized treatment of quantum fluctuations, and merits further study of this system as a platform for \com{exploring complex dynamics of quantum nonlinear systems}.

\section*{Discussion and Outlook}

We have realized a minimal two-mode Kerr system for generating coherent frequency combs under excitation by a single coherent tone. The phase coherence of the generated combs is fundamentally limited by the intrinsic nonlinearity strength in the quantum modes which form the device. The excellent agreement between theory and experiment points toward a highly controllable experimental platform for the study of complex nonlinear dynamics in the quantum regime. Our device realizes a classically-unstable Kerr-nonlinear regime, ideally suited to understand the potentially competing role of strong quantum fluctuations as a source of decoherence and non-classicality in moderate to strongly nonlinear quantum devices.

Finally, the versatility of the cQED platform admits extensions of our device to multimode systems, and to realizations employing tunable parametric couplers~\cite{frattini_3-wave_2017}, paving the way towards an \textit{in-situ} engineerable  multifrequency light source. Such frequency combs could enable multiplexed quantum measurement~\cite{essig_multiplexed_2020} using a single monochromatic incident tone. The generated combs could also function as multifrequency pumps to phase-coherently drive multiple parametric processes simultaneously in a single device for Hamiltonian engineering applications~\cite{sliwa_PRX_2015, Lecocq_PRApplied_2017, metelmann_nonreciprocal_2018, Sivak_PRApplied_2019}. This could include the intriguing possibility of multifrequency pumps exhibiting non-classical coherences, using comb generators operating in the deep-quantum regime.

\section*{ACKNOWLEDGMENTS}
This work was supported by the Charles E. Kaufman Foundation of the Pittsburgh Foundation, by NSF Grant No. PIRE-1743717, and by the Army Research Office under Grant No. W911NF-18-1-0144. The work of S. K. and H. E. T. was additionally supported by the US Department of Energy, Office of Basic Energy Sciences, Division of Materials Sciences and Engineering under Award No. $\text{DE-SC0016011}$. The views and conclusions contained in this document are those of the authors and should not be interpreted as representing the official policies, either expressed or implied, of the Army Research Office or the US Government. The US Government is authorized to reproduce and distribute reprints for government purposes notwithstanding any copyright notation herein.

\appendix


\section{Stochastic description of quantum dynamics via the Positive-$P$ representation}
\label{app:prep}

The derivation of the system Hamiltonian and master equation we consider in this paper is quite standard in cQED; in particular, it may be found in detail in the SI of our previous work~\cite{Saeed2018}, and we thus do not repeat the derivation here. Instead, in this appendix section we begin with the master equation description, derive its corresponding classical description making use of a positive-$P$ phase-space description, and analyze the stability of the resulting system. 

For convenience, we reproduce here the master equation describing the dynamics of the two-mode system:
\begin{align}
    \dot{\hat{\rho}} = -i[\hat{\mathcal{H}},\hat{\rho}] + \kappa \mathcal{D}[\hat{a}]\hat{\rho} + \gamma \mathcal{D}[\hat{b}]\hat{\rho} + \gamma_{\varphi}\mathcal{D}[\hat{b}^{\dagger}\hat{b}]\hat{\rho} 
    \label{appeq:master}
\end{align}
where the system Hamiltonian in the frame rotating with the drive is given by Eq.~(\ref{eq:hsys}) from the main text.

In the weakly nonlinear regime relevant to the experiment, $\Lambda/\kappa \sim O(10^{-2})-O(10^{-3})$, strong driving leads to large mode occupations $\sim O(10^2)-O(10^3)$, rendering standard master equation and even stochastic wavefunction approaches intractable for direct simulation. Such operating regimes are particularly suited to analysis using a phase-space approach to the dynamics of the density operator $\hat{\rho}$. In this appendix section, we describe the approach used in this work, that of the Positive-$P$ representation of the density operator, and the resulting stochastic differential equations (SDEs) it yields. In Section~III of the SI, we also describe how the SDEs may be solved numerically to obtain quantities of interest.

We employ a representation of the density operator in a non-diagonal coherent state basis over both modes $\hat{a}$ and $\hat{b}$:
\begin{align}
    \hat{\rho}(t) &= \int d^2\zeta~P(\vec{\zeta},t)~\hat{\Xi}_{\alpha} \otimes \hat{\Xi}_{\beta} \nonumber \\
    &\equiv \int d^2\zeta~P(\vec{\zeta},t) \cdot \frac{ \ket{\alpha}\bra{\alpha^{\dagger *}} }{e^{\alpha\alpha^{\dagger}} } \otimes \frac{ \ket{\beta}\bra{\beta^{\dagger *}} }{e^{\beta\beta^{\dagger}} }
    \label{appeq:rhoP}
\end{align}
where $\vec{\zeta} = (\alpha,\alpha^{\dagger},\beta,\beta^{\dagger})$ are complex variables describing a classical phase space, $\vec{\zeta} \in \mathbb{C}^4$. For convenience of notation, we use $\zeta_i$ to refer to the $i$th element of the vector $\vec{\zeta}$, for $i = 1,\ldots 4$, and define $d^2\zeta \equiv \prod_i d^2\zeta_i$ as the integration measure over the entire phase space.

Eq.~(\ref{appeq:rhoP}) is simply an expansion of $\hat{\rho}(t)$ in terms of non-diagonal projection operators $\hat{\Xi}_{\alpha}\otimes\hat{\Xi}_{\beta}$, with weights given by the time-dependent function $P(\vec{\zeta},t)$. For the above definition of $\hat{\Xi}_{\alpha}\otimes\hat{\Xi}_{\beta}$, $P(\vec{\zeta},t)$ is positive-definite function that satisfies a Fokker-Planck equation, and therefore may be meaningfully thought of as a classical distribution function; in particular, $P(\vec{\zeta},t)$ is referred to as the Positive-$P$ distribution\cite{drummond_generalised_1980, carmichael_statistical_2002}. 

The above expansion casts the study of the dynamics of $\hat{\rho}(t)$ and operator averages $\avg{\hat{o}} = {\rm tr}\{\hat{o}\hat{\rho}(t)\}$ into an equivalent study of the dynamics of the distribution function $P(\vec{\zeta},t)$ and of probabilistic variables sampled from this distribution function. Phase space approaches therefore first require obtaining the dynamical equation for the distribution function $P(\vec{\zeta},t)$, which takes the form of a nonlinear Fokker-Planck equation:
\begin{align}
\partial_t P(\vec{\zeta},t) = \left( -\partial_i A_{\rm c}^i + \frac{1}{2}\partial_i\partial_j D_{\rm st}^{ij} \right) P(\vec{\zeta},t) 
\label{appeq:fpe}
\end{align}
where $\partial_i \equiv \frac{\partial}{\partial \zeta_i}$ and repeated indices are summed over. Here $A_{\rm cl}^i$ is the $i$th element of the drift vector $\vec{A}_{\rm cl}$ that defines deterministic nonlinear dynamics:
\begin{align}
    \vec{A}_{\rm c} = 
    \begin{pmatrix}
    \left( +i\Delta_{da} - \frac{\kappa}{2} \right)\alpha -i \varg \beta -i\eta  \\
    \left( -i\Delta_{da} - \frac{\kappa}{2} \right)\alpha^{\dagger} +i \varg \beta^{\dagger} + i\eta  \\
    \left( +i\Delta_{db} - \frac{\gamma+\gamma_{\varphi}}{2} \right)\beta + i \Lambda (\beta^{\dagger}\beta)\beta - i \varg \alpha   \\
    \left( -i\Delta_{db} - \frac{\gamma+\gamma_{\varphi}}{2} \right)\beta^{\dagger} - i \Lambda (\beta^{\dagger}\beta)\beta + i \varg \alpha^{\dagger}   
    \end{pmatrix}
\end{align}

On the other hand, $D_{\rm st}^{ij}$ is the $(i,j)$th element of the diffusion matrix $\mathbf{D}_{\rm st}$ that lends `width' to the distribution function. Here it takes the simple form:
\begin{align}
    \mathbf{D}_{\rm st} = 
    \begin{pmatrix}
    \mathbf{0} & \mathbf{0} \\
    \mathbf{0} & \mathbf{D}_{\beta}
    \end{pmatrix},
    \mathbf{D}_{\beta} = 
    \begin{pmatrix}
    (i\Lambda - \gamma_{\varphi})\beta^2 & \gamma_{\varphi}\beta^{\dagger}\beta \\
    \gamma_{\varphi}\beta^{\dagger}\beta & (-i\Lambda - \gamma_{\varphi})(\beta^{\dagger})^2
    \end{pmatrix}
\end{align}
where $\mathbf{0}$ is the 2-by-2 matrix of zeros. Note that the diffusion includes contributions arising from the nonlinearity $\Lambda$ as well as from the the dephasing term $\gamma_{\varphi}$.

In general, the multi-dimensional nonlinear Fokker-Planck equation, Eq.~(\ref{appeq:fpe}), cannot be analytically solved for the distribution function $P(\vec{\zeta},t)$; exceptions include situations where the Fokker-Planck equation is linear or where certain potential conditions are satisfied~\cite{gardiner2009stochastic}. The current system falls under neither category. However, the utility of the Fokker-Planck equation extends beyond the equation itself; one can also obtain a set of equivalent stochastic differential equations (SDEs) describing the dynamics of phase space variables $\vec{\zeta}(t)$ sampled from the Positive-$P$ distribution satisfying the governing Fokker-Planck equation. The set of SDEs takes the form~\cite{carmichael_statistical_2002}:
\begin{align}
    d\vec{\zeta} = \vec{A}_{\rm c}(\vec{\zeta})dt + \sqrt{\Gamma} \mathbf{B}_1(\vec{\zeta}) d\vec{W}_1(t) + \sqrt{\gamma_{\varphi}} \mathbf{B}_2(\vec{\zeta}) d\vec{W}_2(t)
    \label{appeq:sdes}
\end{align}
where $d\vec{W}_i$ are vectors of real, independent Wiener increments. The noise matrices $\mathbf{B}_1$, $\mathbf{B}_2$ are related to the square root of the diffusion matrix, $\mathbf{D}_{\rm st} = \mathbf{B}_{\rm st} \mathbf{B}_{\rm st}^T$, where $\mathbf{B}_{\rm st}~=~\sqrt{\Gamma}\mathbf{B}_1 + \sqrt{\gamma_{\varphi}}\mathbf{B}_2$. They can be written compactly in block form:
\begin{align}
    \mathbf{B}_1  = 
    \begin{pmatrix}
    \mathbf{0} & \mathbf{0}  \\
    \mathbf{b}_1 & \mathbf{0} \\
    \end{pmatrix}
    ,~\mathbf{B}_2 = 
    \begin{pmatrix}
    \mathbf{0} & \mathbf{0}  \\
    \mathbf{0} & \mathbf{b}_2 \\
    \end{pmatrix}
\end{align}
where the 2-by-2 component matrices $\mathbf{b}_1$ and $\mathbf{b}_2$ are given by:
\begin{align}
    \mathbf{b}_1 = 
    \begin{pmatrix}
    e^{i\theta/2}\beta & 0 \\
    0 & e^{-i\theta/2}\beta^{\dagger}
    \end{pmatrix},~
    \mathbf{b}_2 = 
    \sqrt{\frac{\beta^{\dagger}\beta}{2}}
    \begin{pmatrix}
    e^{i\pi/4} & e^{-i\pi/4} \\
    e^{-i\pi/4} & e^{i\pi/4}
    \end{pmatrix}
\end{align}
Finally, we have defined the parameters $\Gamma$ and $\theta$ via:
\begin{align}
    \Gamma e^{i\theta} \equiv i\Lambda -\gamma_{\varphi} \implies \Gamma = \sqrt{\Lambda^2 + \gamma_{\varphi}^2},~\theta = \arctan\left( -\frac{\Lambda}{\gamma_{\varphi}} \right)
\end{align}

Eqs.~(\ref{appeq:sdes}) are the same as Eq.~(\ref{eq:sdes}) from the main text, and are the central equations we employ to analyze the dynamics of the two-mode system.

\section{Classical limit, fixed points, and linear stability}
\label{app:classical}

While Eqs.~(\ref{appeq:sdes}) describe the quantum dynamics of the two-mode system, they also allow us to analyze a well-defined classical limit, where the stochastic terms in Eqs.~(\ref{appeq:sdes}) vanish. Clearly, the dephasing contribution $\propto \sqrt{\gamma_{\varphi}}\mathbf{B}_2$ can be dropped by setting $\gamma_{\varphi} = 0$. However, simply setting taking $\Lambda = 0$ will render the two-mode system linear and eliminate the comb dynamics we are interested in.

Instead, a simple scaling argument allows us to understand the classical limit of the two-mode system. We consider reducing the nonlinearity by a factor $\Lambda \to \Lambda/k$ ($k > 1$), and simultaneously transforming $\vec{\zeta} \to \sqrt{k}\vec{\zeta}$, $\eta \to \sqrt{k}\eta$. Under this transformation, we find that Eqs.~(\ref{appeq:sdes}) become ($\gamma_{\varphi}=0$):
\begin{align}
    d\vec{\zeta} = \vec{A}_{\rm c}(\vec{\zeta})dt + \frac{1}{\sqrt{k}}\sqrt{\Lambda} \mathbf{B}_1(\vec{\zeta}) d\vec{W}_1(t) 
\end{align}
More precisely, the drift vector $\vec{A}_{\rm c}(\vec{\zeta})$ is \textit{invariant} under this transformation, while the stochastic terms are scaled by a factor of $1/\sqrt{k}$. Physically, this transformation indicates that as the strength of the nonlinearity decreases, the deterministic dynamics remain unchanged provided the drive is suitably increased, upto a scaling of the mode amplitudes $\vec{\zeta}$. The stochastic dynamics, on the other hand, are suppressed. The appropriate classical limit that retains nonlinear dynamics can thus be realized by considering weak nonlinearities under sufficiently strong driving. The dynamical equations that describe this classical limit are thus given by:
\begin{align}
    d\vec{\zeta} = \vec{A}_{\rm c}(\vec{\zeta})dt~~({\rm classical~limit,~}k\to\infty)
    \label{appeq:AVecCl}
\end{align}

Upon dropping the stochastic terms, it is clear to see from the now \textit{ordinary} differential equations above (when written out) that $\alpha^{\dagger} = \alpha^*$, $\beta^{\dagger} = \beta^*$; as a result, the deterministic dynamics in the classical limit, Eqs.~(\ref{appeq:AVecCl}), can finally be written down entirely in terms of $\alpha$, $\beta$:
\begin{subequations}
\begin{align}
\dot{\alpha} &= \left( i\Delta_{da} - \frac{\kappa}{2}\right)\alpha - i \varg \beta - i \eta \label{appeq:alphaEq} \\
\dot{\beta} &= \left( i\Delta_{db} - \frac{\gamma}{2}\right)\beta + i \Lambda |\beta|^2\beta - i \varg \alpha \label{appeq:betaEq}
\end{align}
\end{subequations}
For completeness, we note here that the above system is the same as that obtained by writing down the equations of motion for operator averages $\{\avg{\hat{a}}, \avg{\hat{b}}\}$, neglecting correlations (namely performing replacements of the form $\avg{\hat{b}^{\dagger}\hat{b}\hat{b}} \to \avg{\hat{b}^{\dagger}}\avg{\hat{b}}\avg{\hat{b}}$), and finally replacing operator expectation values by complex amplitudes, $\{ \avg{\hat{a}},\avg{\hat{b}} \} \to \{\alpha, \beta\}$; the derivation here provides some context to the approximations underlying this dropping of correlations. 

The linearity of both mode $\hat{a}$ and the coupling $\propto \varg$ enables the linear mode to be integrated out, leading to a single effective dynamical equation for the nonlinear mode amplitude~\cite{Saeed2018}:
\begin{align}
    \dot{\beta} = &\left(i\Delta_{db}-\frac{\gamma}{2} \right)\beta + i \Lambda |\beta|^2\beta -i\varg \chi_a \eta \nonumber \\
    &- \varg^2 \! \int_0^t d\tau~F(\tau)\beta(t-\tau)
    \label{appeq:effNL}
\end{align}
where we have introduced the linear mode susceptibility $\chi_a = (-i\Delta_{da} + \frac{\kappa}{2})^{-1}$, and where the memory kernel for the self-interaction is given by:
\begin{align}
    F(\tau) = e^{(i\Delta_{da} - \kappa/2)\tau}
\end{align}

The classical steady-state of the two-mode system $(\bar{\alpha},\bar{\beta})$ may be obtained by setting $\dot{\bar{\beta}} = 0$ in Eq.~(\ref{appeq:effNL}). This requirement simplifies the self-interaction term and is exactly equivalent to performing a Markov regime reduction of the same. The result is a cubic polynomial in $|\bar{\beta}|^2$ that can be solved exactly for the steady-state nonlinear mode amplitude $\bar{\beta}$:
\begin{align}
  \left[ \left( \widetilde{\Delta}_{db} + \Lambda|\bar{\beta}|^2 \right)^2  + \frac{\widetilde{\gamma}^2}{4} \right] |\bar{\beta}|^2 = \varg^2 |\chi_a|^2 \eta^2 
    \label{appeq:ssb}
\end{align}
where we have introduced the renormalized nonlinear mode detuning and damping parameters respectively:
\begin{align}
    \widetilde{\Delta}_{db} &= \omega_d - (\omega_b + \varg^2 |\chi_a|^2 \Delta_{da}) \nonumber \\
    \widetilde{\gamma} &= \gamma + \varg^2 |\chi_a|^2 \kappa
\end{align}

The steady-state linear mode amplitude may then be determined by requiring $\dot{\bar{\alpha}} = 0$ in Eq.~(\ref{appeq:alphaEq}), which simply relates $\bar{\alpha}$ to $\bar{\beta}$:
\begin{align}
    \bar{\alpha} =  - \chi_a \left( i \varg \bar{\beta} + i \eta \right)
    \label{appeq:ssa}
\end{align}

Once the steady-state amplitudes $(\bar{\alpha},\bar{\beta})$ have been determined, we perform a stability analysis for small fluctuations around these steady-state(s). Formally, such an analysis can be performed on the linearized version of the effective nonlinear mode dynamical equation, which can be studied analytically \textit{exactly} in the Laplace domain, and is particularly tractable for the special case where $\Delta_{da} = 0$. Full details of such an analysis are provided in Ref.~\cite{Saeed2018}.

However, the current experiment explores more general operating conditions where $\Delta_{da} \neq 0$ in general. In this case, it proves most convenient to simply perform a numerical stability analysis based on the Jacobian matrix of the original two-mode system. Performing the linearized stability analysis requires expanding Eqs.~(\ref{appeq:AVecCl}) around the classical steady state $(\bar{\alpha},\bar{\beta})$. For notational convenience, we define the vector of steady-state amplitudes $\vec{Z}$ and small fluctuations $\vec{z}(t)$ respectively:
\begin{subequations}
\begin{align}
    \vec{Z} &= (\bar{\alpha},\bar{\alpha}^*,\bar{\beta},\bar{\beta}^*)^T  \\
    \vec{z}(t) &= (\delta\alpha(t),\delta\alpha^*(t),\delta\beta(t),\delta\beta^*(t))^T
    \label{appeq:ZDef}
\end{align}
\end{subequations}

\begin{figure}[t]
 	\includegraphics[scale = 1.0]{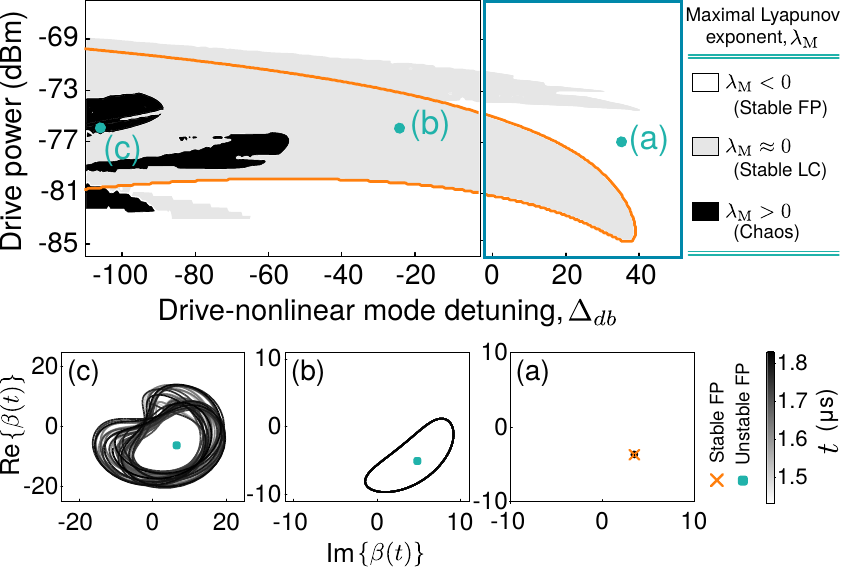}
 	\caption{\textbf{Calculated maximal Lyapunov exponent $\lambda_{\rm M}$ in Drive-$\Delta_{db}$ space.} The panel on the right framed in blue indicates the detuning range explored in Figs. 2, 3 of the main text. Solid orange curve indicates the linear stability boundary. In the white regions, $\lambda_{\rm M} < 0$ is negative and the system is therefore stable. In the light gray regions, $\lambda_{\rm M} \approx 0$, indicating a stable limit cycle. The dark regions are where $\lambda_{\rm M} > 0$, and the system exhibits chaotic dynamics. Typical long-time dynamics projected in the nonlinear mode phase space are plotted in (a)-(c) corresponding to dynamics in the stable fixed point, stable limit cycle, and chaotic regimes respectively.}
 	\label{fig:lyapunov}
 \end{figure}

Then, we expand the variables $\vec{\zeta}(t)$ around the steady-state $\vec{Z}$:
\begin{align}
    \vec{\zeta}(t) = \vec{Z} + \vec{z}(t)
\end{align}
and linearize Eqs.~(\ref{appeq:AVecCl}) in small fluctuations $\vec{z}(t)$, obtaining the set of equations:
\begin{align}
    \frac{d\vec{z}}{dt} = \mathbf{J}[\vec{Z}] \cdot \vec{z}(t)
    \label{eq:linEq}
\end{align}
where $\mathbf{J}[\vec{Z}]$ defines the Jacobian matrix of the two-mode system evaluated at the classical steady-state; its entries are given by $J_{ij} = \partial_j A_{\rm c}^i$, where $A_{\rm c}^i$ is the $i$th element of $\vec{A}_{\rm c}$; more explicitly the Jacobian matrix takes the form:
\begin{widetext}
\begin{align}
    \mathbf{J}[\vec{Z}] = 
    \begin{pmatrix}
    +i\Delta_{da} - \frac{\kappa}{2} & 0 & -i\varg & 0 \\
    0 & -i\Delta_{da} - \frac{\kappa}{2} & 0 & i\varg \\
    -i\varg & 0 & +i\Delta_{db} - \frac{\gamma}{2} + i 2\Lambda |\bar{\beta}|^2 & i\Lambda (\bar{\beta}^2) \\
    0 & i\varg & -i\Lambda (\bar{\beta}^*)^2 & -i\Delta_{db} - \frac{\gamma}{2} - i 2\Lambda |\bar{\beta}|^2
    \end{pmatrix}
    \label{eq:ssJac}
\end{align}
\end{widetext}
The stability of Eqs.~(\ref{eq:linEq}) is determined by the eigenvalues of the above Jacobian matrix, obtained by setting $\det \mathbf{J} = 0$; these are used to determine the stability boundaries obtained in the main text, and in Fig.~\ref{fig:lyapunov} of the following appendix section.


\section{Numerical phase diagram and Lyapunov stability}
\label{app:lyapunov}
 
Regions in the classical phase diagram with no stable fixed points, Fig.~\ref{fig:varyingWQ}, can give rise to a rich class of dynamics. Amongst various metrics to characterize such dynamics, we employ the standard technique of computing the maximal Lyapunov exponent $\lambda_{\rm M}$, which describes the sensitivity of dynamical trajectories to small perturbations in the long-time limit. For details of the numerical approach tom computing $\lambda_{\rm M}$, see SI~\cite{SI}. 

The maximal Lyapunov exponent $\lambda_{\rm M}$ we calculate is plotted for Device A parameters in drive-$\Delta_{db}$ space in Fig.~\ref{fig:lyapunov}; the panel on the right framed in blue shows the region of drive-$\Delta_{db}$ space explored in Fig.~\ref{fig:varyingWQ}. The blank regions indicate regions where $\lambda_{\rm M} < 0$, indicating a stable fixed point; perturbations near this point decay over time, settling back towards the fixed point. This is visible in the projection of the steady-state dynamics onto the nonlinear mode phase space, plotted in Fig.~\ref{fig:lyapunov}~(a); in the long time limit the system has returned to the stable fixed point indicated by the orange cross. The gray regions indicate $\lambda_{\rm M} \approx 0$, signifying a stable limit cycle attractor~\cite{haken_at_1983}. Steady-state dynamics here follow a stable phase space orbit, as shown in Fig.~\ref{fig:lyapunov}~(b), around a classically unstable fixed point (green square). The periodic orbits yield combs in the frequency domain, as observed in Fig.~\ref{fig:lyapunov}.

Finally, the dark regions indicate $\lambda_{\rm M} > 0$. Here perturbations grow without bound over time, manifesting in dynamical chaos observed in numerical simulations of the classical system. The steady-state dynamics plotted in Fig.~\ref{fig:lyapunov}~(c) show how over time a single fixed orbit does not emerge and the system explores a large region of phase space in an irregular manner. The region framed in blue in the phase diagram describes the detuning range explored in the experiment, Figs.~2, 3 of the main text, where the system exhibits stable limit cycle dynamics, consistent with observations in the main text. However, for much more negative $\Delta_{db}$ it is possible to observe chaos with the same system. This indicates the potential of the two-mode system for controlled studies of chaos in the quantum regime; hints of this dynamics are seen in Fig.~4 of the main text, as well as for Device B (see SI).

\section{Quantum simulations: comb coherence and estimating pure dephasing rate $\gamma_{\varphi}$}
\label{app:nlDephasing}

Simulating Eqs.~(\ref{eq:sdes}) allows us to calculating the output coherence function, Eq.~(\ref{eq:g1}) defined in the main text; technical details of the simulations are included in Section III of the SI~\cite{SI}. This enables us to extract the coherence time $T_{\rm coh}$, as discussed in the main text. The only parameter required to simulate the SDEs that we are unable to directly measure is the pure dephasing rate $\gamma_{\varphi}$; the weak nonlinearity of the nonlinear mode prevents standard Ramsay measurement of the pure dephasing rate, and indirect methods based on cavity measurement are limited by the large disparity between the dephasing rate and the cavity linewidth $\kappa$. These difficulties are discussed in Section IV of the SI~\cite{SI}.

\begin{figure}[t]
 	\includegraphics[scale = 1.0]{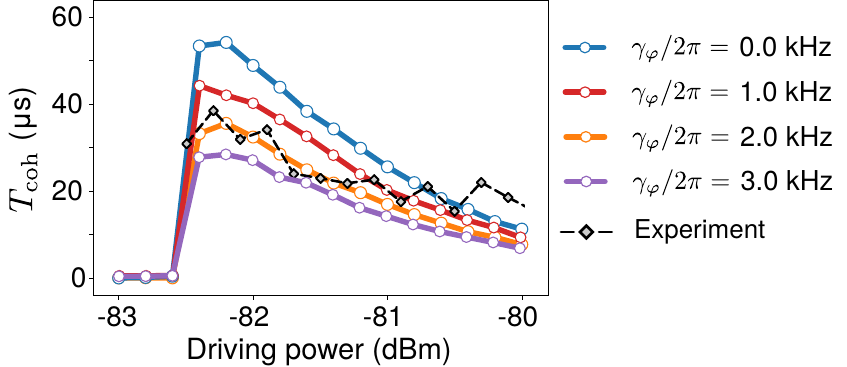}
 	\caption{\textbf{Coherence times as a function of $\gamma_{\varphi}$.} Colored lines and points show numerically obtained $T_{\rm coh}$ values from the simulation of Eqs.~(\ref{eq:sdes}) as a function of drive power, while dashed-diamonds indicate experimental values. }
 	\label{fig:gammaPhiFit}
 \end{figure}

However, the coherence of frequency combs is affected by the known nonlinearity and the unknown pure dephasing rate; as a result, by simulating Eqs.~(\ref{eq:sdes}) for various values of $\gamma_{\varphi}$ and comparing with experimental observations, we can estimate $\gamma_{\varphi}$. In Fig.~\ref{fig:gammaPhiFit}, we show the numerically obtained value of $T_{\rm coh}$ across the same cross-section of the phase diagram included in the main text, Fig.~2(b), for $\gamma_{\varphi}/(2\pi) \in [0.0,1.0,2.0,3.0]~\text{kHz}$. Also shown is the experimental result. From these results we conclude that the pure dephasing rate may be well approximated to lie within $\gamma_{\varphi}/(2\pi) \in [1.0,3.0]~\text{kHz}$. Furthermore, the best fit is found to be for $\gamma_{\varphi}/(2\pi) \simeq 2.0~\text{kHz}$.

\section{Linearized Floquet Analysis of SDEs}
\label{app:floquet}

The influence of quantum noise on system dynamics as described by the stochastic terms in Eqs.~(\ref{eq:sdes}) is well understood when considering dynamics near a classically stable fixed point. Here one linearizes the system around the stable fixed point and studies weak fluctuations due to stochastic terms. However, in the frequency comb regime the system exhibits no classically stable fixed points, instead settling into a stable attractor describing a limit cycle. The study of linearized fluctuations around such stable attractors has gained much interest recently and can be performed by linearizing the dynamics around the periodic classical solution~\cite{demir_phase_2000, navarrete-benlloch_general_2017}. 

To begin, we rewrite the system of SDEs, Eqs.~(\ref{appeq:sdes}), in the form below:
\begin{align}
    \frac{d \vec{\zeta}}{dt} = \vec{A}_{\rm c}(\vec{\zeta})  + \mathbf{B}_{\rm st}(\vec{\zeta}) \cdot\frac{d\vec{W}(t)}{dt}  
\end{align}
where we have suppressed the dependence of $\vec{A}_{\rm c}, \mathbf{B}_{\rm st}$ on system parameters for notational convenience. In the frequency comb regime, the classical (noise-free) system admits the periodic solution $\vec{\zeta}_{\rm c}(t)$:
\begin{align}
    \frac{d\vec{\zeta}_{\rm c}}{dt} = \vec{A}_{\rm c}(\vec{\zeta}_{\rm c}) 
    \label{eq:Z}
\end{align}
For frequency combs with spacing $\Delta$, $\vec{\zeta}_{\rm c}(t)$ is periodic with period $T = \frac{2\pi}{\Delta}$.

We can then consider fluctuations $\vec{z}(t)$ around this classical periodic solution:
\begin{align}
    \vec{\zeta}(t+\theta) = \vec{\zeta}_{\rm c}(t+\theta) + \vec{z}(t+\theta)
    \label{eq:zExp}
\end{align}
where we have introduced the additional phase parameter $\theta(t)$ which is not fixed by the classical dynamical equations of motion, and is therefore susceptible to perturbations due to noise (or other external stimuli)~\cite{demir_phase_2000, navarrete-benlloch_general_2017}. We are now interested in the linearized dynamics of the fluctuations $\vec{z}(t+\theta)$. Substituting the expansion, Eq.~(\ref{eq:zExp}), into the system of SDEs, Eq.~(\ref{eq:sdes}), and retaining only terms linear in $\vec{z}(t)$, we find:
\begin{align}
    \frac{d\vec{z}}{dt} + \frac{d\vec{\zeta}_{\rm c}}{dt}\dot{\theta} = \mathbf{J}[\vec{\zeta}_{\rm c}(t)]\cdot\vec{z} + \mathbf{B}_{\rm st}[\vec{\zeta}_{\rm c}(t)]\cdot\frac{d\vec{W}}{dt}
    \label{appeq:linSDEs}
\end{align}
where $\mathbf{J}[\vec{\zeta}_{\rm c}(t)]$ is the Jacobian matrix evaluated along the \textit{periodic} classical solution, and is therefore a periodic matrix itself. Similarly $\mathbf{B}_{\rm st}[\vec{\zeta}_{\rm c}(t)]$ is the noise matrix also evaluated along the periodic classical solution. Finally, $\frac{d\vec{\zeta}_{\rm c}}{dt} \equiv \vec{v}$ is the velocity vector and is tangential to the limit cycle trajectory. This term clearly vanishes if $\vec{\zeta}_{\rm c}(t)$ is time \textit{independent}, as in the case of a stable fixed point where $\vec{\zeta}_{\rm c}(t) \to \vec{Z}$ defined in Eq.~(\ref{appeq:ZDef}); then the above equation simply describes the linearized dynamics of fluctuations around the fixed point, governed by a static Jacobian and driven by noise terms. 

Here, however, the velocity term does not vanish and in addition to the dynamics of $\vec{z}(t)$, we are also interested in the evolution of the free phase $\theta(t)$ under the influence of stochastic terms. To solve for the dynamics of a system governed by a time-periodic dynamical matrix, it proves useful to express the linearized fluctuations $\vec{z}(t)$ in terms of the Floquet eigenvectors defined by the \textit{linearized} periodic system:
\begin{align}
    \frac{d\vec{z}}{dt} = \mathbf{J}[\vec{\zeta}_{\rm c}(t)]\cdot\vec{z} 
    \label{appeq:linFloquet}
\end{align}
Details of the Floquet eigensystem analysis are provided in the SI~\cite{SI}; here for clarity we restrict our discussion to understanding how the main results can be used to analyze limit cycle phase diffusion. For convenience we define the periodic dynamical matrix $\mathbf{J}[\vec{\zeta}_{\rm c}(t)] \equiv \mathbf{J}(t)$ and the periodic noise matrix $\mathbf{B}_{\rm st}[\vec{\zeta}_{\rm c}(t)] \equiv \mathbf{B}_{\rm st}(t)$. The Floquet eigenvectors $\{\vec{p}_i(t),\vec{q}_i(t)\}$ for $i = 0,\ldots,N-1$ where $N$ is the dimension of the system of ODEs ($N=4$ for the present system), are periodic with the period of the stable classical limit cycle, $T$. They themselves satisfy the linear systems of equations:
\begin{subequations}
\begin{align}
    \dot{\vec{p}}_i(t) &= \left[\mathbf{J}(t)-\mu_i \right]\vec{p}_i(t) \label{appeq:pdot} \\
    \dot{\vec{q}}^{\dagger}_i(t) &= \vec{q}_i^{\dagger}(t)\left[\mu_i-\mathbf{J}(t) \right] \label{appeq:qdot}
\end{align}
\end{subequations}

The $\{\mu_i\}$ are Floquet exponents determined by the eigenvalues of the fundamental matrix of the Floquet system. For systems with a periodic stable attractor, at least one of the Floquet exponents, which we label $\mu_0$ here, vanishes~\cite{haken_at_1983}. The corresponding Floquet eigenvector $\vec{p}_0(t)$ can be shown to be proportional to the tangential velocity vector $\vec{v}$ (see SI~\cite{SI}). Finally, the Floquet eigenvectors satisfy the following orthogonality relation:
\begin{align}
    \vec{q}_j^{\dagger}(t)\vec{p}_i(t) = \delta_{ij}~\forall~t \in [0,T]
\end{align}

To proceed, we expand the weak fluctuations around the stable limit cycle in terms of the Floquet eigenvectors:
\begin{align}
    \vec{z}(t) = \sum_{n=1}^{N-1} c_n(t) \vec{p}_n(t)
\end{align}
Note that the above expansion does not include the Floquet eigenvector $\vec{p}_0(t)$ corresponding to $\mu_0 = 0$, which as mentioned before is proportional to the tangent vector to the classical limit cycle~\cite{navarrete-benlloch_general_2017}. Substituting the above expansion into the linearized set of SDEs, Eqs.~(\ref{appeq:linSDEs}), we find:
\begin{align}
    &\sum_{n=1}^{N-1} \left[\dot{c}_n(t)\vec{p}_n(t) + c_n(t)\dot{\vec{p}}_n(t)\right] + \vec{v}\dot{\theta} \nonumber \\
    &= \mathbf{J}(t)\cdot \sum_{n=1}^{N-1} c_n(t) \vec{p}_n(t) + \mathbf{B}_{\rm st}(t)\cdot \frac{d\vec{W}}{dt}
\end{align}
where we now use Eq.~(\ref{appeq:pdot}) to eliminate $\mathbf{J}(t)\vec{p}_n(t)$; the terms corresponding to time derivatives of the right Floquet eigenvectors simply cancel, and we finally obtain:
\begin{align}
\sum_{n=1}^{N-1} \dot{c}_n(t)\vec{p}_n(t)  + \vec{v}\dot{\theta} = \sum_{n=1}^{N-1} \mu_n c_n(t)\vec{p}_n(t) + \mathbf{B}_{\rm st}(t)\cdot \frac{d\vec{W}}{dt} 
\end{align}
The remaining terms can be used to obtain equations of motion for the expansion coefficients. However, we are primarily interested in the diffusion of the phase variable $\theta(t)$. We can use the fact that $\vec{v}(t) \propto \vec{p}_0(t)$ to isolate the equation of motion for the phase variable: multiplying by the Floquet left eigenvector $\vec{q}_0^{\dagger}(t)$ and using the orthogonality of the Floquet eigenvectors, the above system simplifies to:
\begin{align}
    \left(\vec{q}_0^{\dagger}(t)\vec{v}(t) \right) \dot{\theta}(t) = \vec{q}_0^{\dagger}(t) \left( \mathbf{B}_{\rm st}(t)\cdot \frac{d\vec{W}}{dt} \right)
\end{align}
For notational simplicity, we can normalize $\vec{q}_0(t)$ (and therefore $\vec{p}_0(t)$) such that $\vec{q}_0^{\dagger}(t)\vec{v} = v_T$ where $v_T$ is the root-mean-square velocity over the limit cycle period $T$, $v_T=\sqrt{\frac{1}{T}\int_0^T dt~||\vec{v}(t)||^2}$. Then, defining the time dependent noise projection of the noise vector in parenthesis onto $\vec{q}_0(t)$:
\begin{align}
    n(t) =  \vec{q}_0^{\dagger}(t) \left( \mathbf{B}_{\rm st}(t)\cdot \frac{d\vec{W}}{dt} \right)
\end{align}
we obtain the dynamical equation for $\theta(t)$:
\begin{align}
    v_T \dot{\theta}(t) = n(t)
\end{align}
which is the equation introduced in the main text. However, note that as introduced, the phase variable is a perturbation to the time $t$; it appears in the comb time evolution multiplied by the relevant frequency scale for frequency comb, namely the comb spacing $\Delta$. We then have the equation of motion:
\begin{align}
    r_{\rm eff} \left[\Delta\dot{\theta}(t)\right] = n(t)
\end{align}
where we introduce the effective limit cycle radius $r_{\rm eff}$ via $v_T = r_{\rm eff} \Delta$ as in the main text.
The simplified notation does require some caution; the noise term $n(t)$ is a stochastic term and solutions to the above equation must ultimately be determined by calculating moments of the phase variable. In particular, we can obtain the variance:
\begin{align}
    \Delta^2\avg{\theta^2(t)} = \frac{1}{r_{\rm eff}^2}\int_0^t \int_0^t d\tau~d\tau'~\avg{n(\tau) n(\tau')}
\end{align}
where we have set $\theta(0) = 0$, since only the relative phase is important. The double integral above simplifies once the noise correlation functions for white noise variables $\frac{d\vec{W}}{dt} = \vec{\xi}(t)$ are substituted. In particular since:
\begin{align}
    \avg{\xi_i(\tau)\xi_j(\tau')} = \delta_{ij}\delta(\tau-\tau')
\end{align}
we can write the variance of noise moments as:
\begin{align}
    \avg{n(\tau) n(\tau')} \equiv \avg{n(\tau)^2}\delta(\tau-\tau')
\end{align}
With the above definition, we can write the phase variance after a time equal to the period $T$ as:
\begin{align}
    \Delta^2\avg{\theta^2(T)} = \frac{1}{r_{\rm eff}^2}\int_0^T d\tau\avg{n(\tau)^2} = T\left(\frac{\delta n}{r_{\rm eff}}\right)^2 \equiv  2\left(\frac{T}{T_{\rm coh}}\right)
\end{align}
where we have introduced the average projected noise standard deviation as in the main text, $\delta n = \sqrt{\frac{1}{T}\int_0^T d\tau~\avg{n(\tau)^2}}$. The inset of the phase diagram in Fig. 3 of the main text plots $T_{\rm coh}$ as the limit cycle coherence time.

\bibliography{refs}

\clearpage

\newpage
\newpage

\onecolumngrid

\begin{center}
{\large \textbf{Supplementary Material for ``Nearly quantum-limited Josephson-junction Frequency Comb synthesizer"} } \\

\vspace{\baselineskip}

\flushleft{
Pinlei Lu*, Tzu-Chiao Chien, Xi Cao, Olivia Lanes, Chao Zhou, and Michael J. Hatridge \\
{\small\textit{Department of Physics and Astronomy, University of Pittsburgh}} \\
S. Khan*, H. E. T\"ureci \\
{\small
\textit{Department of Electrical Engineering, Princeton University} \\
(Dated: May 18, 2020) } \\
{\small
\textit{*These authors contributed equally to this publication.}}}
\end{center}

\vspace{\baselineskip}


\setcounter{equation}{0}

\setcounter{page}{1}
\thispagestyle{empty}

\begin{center}
    \rule{10cm}{1pt}
\end{center}

\section*{Lyapunov stability}
\label{sec:lyapunov}

In this section we describe a standard numerical approach to the calculation of Lyapunov exponents of dynamical systems, which we use for the analysis in Appendix~C of the manuscript. Employing the notation from that Appendix, consider a deterministic trajectory $\vec{\zeta}(t)$ evolving according to the nonlinear equations of motion describing the classical system, Eqs.~(B2). Linearized fluctuations about this trajectory, $\vec{z}(t)$, are then propagated by the time-dependent Jacobian matrix evaluated along the deterministic trajectory:
\begin{align}
    \frac{d\vec{z}}{dt} = \mathbf{J}[\vec{\zeta}(t)] \cdot \vec{z}(t)
    \label{eq:lyEq}
\end{align}
which is simply the system given by Eqs.~(B11) from the manuscript, Appendix~B, but with the Jacobian now evaluated along a general time-dependent trajectory $\vec{\zeta}(t)$. Such trajectories may exhibit complicated dynamics, but being governed by a linear system, this evolution is ultimately always related to characteristic exponents obtained from the dynamical matrix. The maximal Lyapunov exponent $\lambda_{\rm M}$ is the largest such exponent (accounting for sign, not magnitude); it plays the role that the largest eigenvalue would play in the case of a dynamical system governed by a time-\textit{independent} dynamical matrix. The maximal Lyapunov exponent may be computed by studying the long-time dynamics of trajectories governed by Eq.~(\ref{eq:lyEq}):
\begin{align}
    \lambda_{\rm M} = \lim_{t\to \infty} \frac{1}{t} \log \frac{||\vec{z}(t)||}{||\vec{z}(0)||}
    \label{eq:maxL}
\end{align}

However, since the evolution of trajectories is governed by a linear dynamical equation, these trajectories may grow unbounded exponentially with time if the Lyapunov exponent is positive. In practice this exponential growth renders the above expression intractable for numerical computations. An alternative procedure to circumvent this issue begins by separating the time evolution from $[0,t]$ into a series of $N_p$ consecutive short-time intervals, $[\tau_0,\tau_1,\ldots,\tau_{N_p}]$, where $\tau_0 = 0, \tau_{N_p} = t$, and $\tau_p - \tau_{p-1} = \Delta\tau$ is the short-time interval spacing. We then solve for $\vec{z}^{(p)}(\tau)$, $\tau \in [\tau_{p-1},\tau_p]$ with $p = 1,\ldots, N_p$, obtaining $N_p$ individual trajectory vectors $\{\vec{z}^{(p)}(\tau)\}$. Such an evolution would be identical to the entire evolution over $[0,t]$ \textit{if} we imposed $\vec{z}^{(p)}(\tau_{p-1}) = \vec{z}^{(p-1)}(\tau_{p-1})$, requiring continuity of the solution at the endpoints of each time interval. Consequently, the process would do nothing to alleviate the problem of unbounded growth. To guard against the latter, we additionally require the initial trajectory at the beginning of every evolution interval to be normalized to one:
\begin{align}
    \vec{z}^{(p)}(\tau_{p-1}) = \frac{\vec{z}^{(p-1)}(\tau_{p-1})}{||\vec{z}^{(p-1)}(\tau_{p-1})||} \implies ||\vec{z}^{(p)}(\tau_{p-1})|| = 1
    \label{eq:iterateNorm}
\end{align}
Then, for the $p$th iterate, we can estimate the maximal Lyapunov exponent as:
\begin{align}
    \lambda_{\rm M}^{(p)} = \frac{1}{\Delta\tau} \log \frac{||\vec{z}^{(p)}(\tau_{p})||}{||\vec{z}^{(p)}(\tau_{p-1})||} = \frac{1}{\Delta\tau} \log ||\vec{z}^{(p)}(\tau_{p})||
\end{align}
which simply measures the growth of the norm of the $p$th trajectory $\vec{z}^{(p)}(\tau)$ for $\tau \in [\tau_{p-1},\tau_p]$. 

An estimate for the actual maximal Lyapunov exponent is then finally computed by averaging over the iterations:
\begin{align}
    \lambda_{\rm M} \approx \frac{1}{N_p} \sum_{p=1}^{N_p} \lambda_{\rm M}^{(p)} 
\end{align}

The above procedure ensures a faithful and \textit{bounded} evolution of a single trajectory over a time $N_p \Delta\tau$, \textit{provided} $\Delta\tau$ is chosen judiciously relative to $|1/\lambda_{\rm M}|$; the latter sets the characteristic evolution timescale for the trajectories and is of course \textit{a priori} unknown. If $\Delta\tau \gg |1/\lambda_{\rm M}|$, trajectories may grow too substantially during the evolution if $\lambda_{\rm M} > 0$, leading to the same numerical errors present in the original formulation, Eq.~(\ref{eq:maxL}). If instead $\Delta\tau \ll |1/\lambda_{\rm M}|$, the normalized trajectories will remain mostly unchanged from their initial values, leading to an estimate of $\lambda_{\rm M}^{(p)} \to 0$ regardless of its actual value. Therefore an intermediate $\Delta\tau$ value must be employed, keeping in mind also that smaller $\Delta\tau$ values necessarily require the use of more iterations $N_p$ for convergence. In practice, the convergence of the estimate may be evaluated by calculating $\lambda_{\rm M}$ as a function of increasing $N_p$ for a fixed $\Delta\tau$, until $\lambda_{\rm M}$ remains approximately unchanged with further iterations. Comparing a set of such estimates for a range of $\Delta\tau$ values then ensures that a consistent estimate is obtained.

The maximal Lyapunov exponent $\lambda_{\rm M}$ obtained with this approach is plotted for Device A parameters in drive-$\Delta_{db}$ space in Fig.~5 of the manuscript.

\section*{Supplementary experimental details and results}
\label{sec:suppExpt}

In this section, we include details of additional experimental measurements, including the measured phase diagram for Device B, as well as the measurement of the Kerr nonlinearity strength.

\subsection*{Phase diagram for Device B (5 SQUIDs)}
\label{ssec:5JJ}


\begin{figure*}[t]
 	\includegraphics[scale = 0.65]{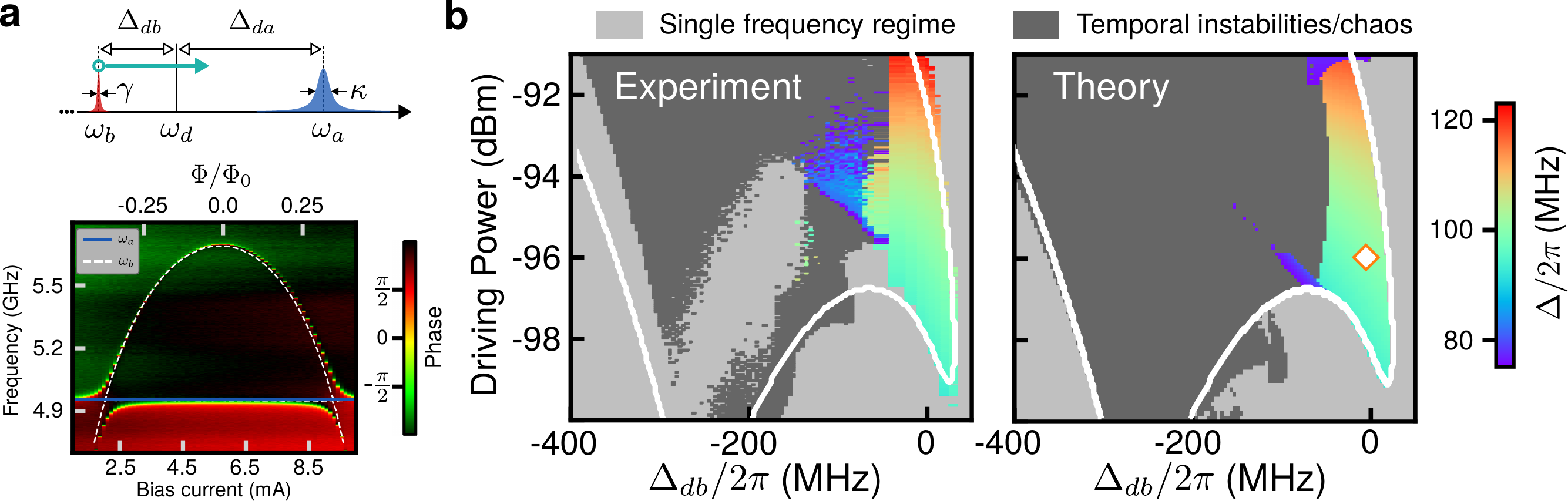}
 	\caption{\textbf{Device B measurements.} (a) Flux sweep showing the system polariton modes and the avoided crossing as the nonlinear mode is swept across the linear mode resonance. The horizontal blue line indicates the bare linear mode frequency, while the white dashed line indicates the bare nonlinear mode frequency. (b) Experimental and theoretical phase diagram in Drive-$\Delta_{db}$ space, as the nonlinear mode frequency is swept (see schematic in (a)). The observed comb spacing $\Delta$ in the multifrequency regime is plotted, alongside the single frequency regime (light gray) and regime with temporal instabilities (dark gray). White curve in both phase diagrams is the analytically obtained boundary enclosing the classically unstable region. Orange diamond indicates the drive power (-96 dBm) and detuning ($\Delta_{db}/2\pi = -5.87~$MHz) for which coherence function results are plotted in Fig.~3(c) of the main text for Device B.}
 	\label{fig:5JJ}
 \end{figure*}


In addition to Device A, which employs a 25 SQUID array, we explore the impact of nonlinearity on comb dynamics by fabricating Device B, which employs a 5 SQUID array and therefore possesses an approximately 25-fold stronger nonlinearity. In Fig.~\ref{fig:5JJ}~(a), we show the flux sweep of this device, indicating the polariton resonances of the two-mode system. Fitting to the avoided crossing reveals a coupling strength of $\varg/2\pi = 89.25$~MHz, similar to Device A (by design), and a bare linear mode linewidth of $\kappa/2\pi = 22.84$~MHz. 

Having verified that the device satisfies the strong coupling condition $\varg > \frac{\kappa}{2}$ at resonant driving ($\Delta_{da} = 0$, see main text), we can explore the classically predicted unstable regime as was done for Device A. Fixing the driving frequency at $\omega_d/2\pi = 4.9085~$GHz, we change the external flux through the SQUIDs to sweep the nonlinear mode frequency, as shown schematically in the top panel of Fig.~\ref{fig:5JJ}~(a). Device B has a much larger flux modulation range than Device A, enabling us to explore a wider range of drive-nonlinear mode detunings $\Delta_{db}$. The resulting phase diagram in drive power-$\Delta_{db}$ space is shown in Fig.~\ref{fig:5JJ}~(b), with the theoretically predicted phase diagram shown in the right panel, both plotted with the same axes. The light gray regions indicate the single frequency regime, which gives way to a multifrequency comb regime at appropriate drive strengths for small $|\Delta_{db}|$. The experiment and theory agree quite well both in terms of the critical detuning where the combs emerge, as well as the observed comb spacings. Finally the orange diamond in the theory plot indicates the position on the phase diagram for which coherence function results are plotted in Fig.~3 of the main text.

Note that for more negative $\Delta_{db}$, a region (dark gray) emerges where the system exhibits temporal instabilities similar to Device A, in both the experimental and theoretical phase diagrams. Numerical simulations here indicate that the system exhibits chaotic dynamics (maximal Lyapunov exponent $\lambda_{\rm M} > 0$). We also find greater disparity between experiment and theory here; in addition to possible deviations from classical predictions due to quantum effects, the dynamics exhibit temporal instabilities that require careful processing. Such dynamical regimes therefore merit further detailed investigation. The white contour in both figures depicts the analytically predicted unstable region, as determined by the linearized dynamics of Eqs.~(B11) of the manuscript; it agrees well with both experiment and numerical simulations, in particular for small $|\Delta_{db}|$. 

\subsection*{Kerr nonlinearity measurement}
\label{ssec:kerr}

To demonstrate the dependence of comb coherence on the quantum nature of the device nonlinearity, knowledge of this engineered Kerr nonlinearity strength is of crucial importance. Typically, one would do so via a standard pump-probe measurement that measures the Kerr-induced frequency shift of the nonlinear mode as the pump power incident on it increases. However, for the two-mode system such a measurement accesses the frequency shift of the renormalized \textit{polariton} modes of the system, which of course depends on the degree of hybridization between linear and nonlinear modes. In this section we clarify how measured polariton mode frequency shifts can be used to extract the bare nonlinear mode Kerr interaction strength.

We begin by rewriting below, for convenience, the full system Hamiltonian ($\hbar=1$) from Eq.~(1) of the main text:
\begin{align}
    \hat{\mathcal{H}} = -\Delta_{da}\hat{a}^{\dagger}\hat{a} - \Delta_{db} \hat{b}^{\dagger}\hat{b} -\frac{\Lambda}{2} \hat{b}^{\dagger}\hat{b}^{\dagger}\hat{b}\hat{b} + \varg (\hat{a}^{\dagger}\hat{b} + \hat{a}\hat{b}^{\dagger} ) + \eta (\hat{a} + \hat{a}^{\dagger} )
    \label{eq:hsys}
\end{align}
Next, we consider the linear Hamiltonian $\hat{\mathcal{H}}_{\rm L}$ that determines the polariton modes:
\begin{align}
    \hat{\mathcal{H}}_{\rm L} = \omega_{a}\hat{a}^{\dagger}\hat{a} + \omega_{b}\hat{b}^{\dagger}\hat{b} + \varg(\hat{a}^{\dagger}\hat{b} + \hat{a}\hat{b}^{\dagger} ) \equiv 
    \begin{pmatrix}
    \hat{a}^{\dagger} & \hat{b}^{\dagger} \\
    \end{pmatrix}
    \underbrace{
    \begin{pmatrix}
    \omega_{a} & \varg \\
    \varg & \omega_{b}
    \end{pmatrix}
    }_{\mathbf{H}_{\rm L}}
    \begin{pmatrix}
    \hat{a} \\ 
    \hat{b}
    \end{pmatrix}
\end{align}
which is obtained from Eq.~(\ref{eq:hsys}) by neglecting the nonlinearity and drive terms, and returning to the lab frame. The above Hamiltonian may be diagonalized by introducing the matrix of eigenvectors $\mathbf{P}$ and diagonal matrix of eigenvalues $\mathbf{D}$ for the matrix $\mathbf{H}_{\rm L}$, such that $\mathbf{H}_{\rm L} = \mathbf{P}\mathbf{D}\mathbf{P}^{-1}$. The Hamiltonian then becomes:
\begin{align}
    \hat{\mathcal{H}}_{\rm L} = \nu_a \hat{c}_a^{\dagger}\hat{c}^{\ }_a + \nu_b \hat{c}_b^{\dagger}\hat{c}_b^{\ },~
    \begin{pmatrix}
    \hat{c}^{\ }_a \\
    \hat{c}^{\ }_b 
    \end{pmatrix}
    = \mathbf{P}^{-1}
    \begin{pmatrix}
    \hat{a} \\ 
    \hat{b}
    \end{pmatrix},~
    \mathbf{D} = 
    \begin{pmatrix}
    \nu_a & 0 \\
    0 & \nu_b
    \end{pmatrix}
    \label{eq:polDef}
\end{align}
which serves to define the polariton modes $\hat{c}_a$, $\hat{c}_b$, and corresponding frequencies $\nu_a$, $\nu_b$. 


\begin{figure*}[t]
 	\includegraphics[scale = 0.32]{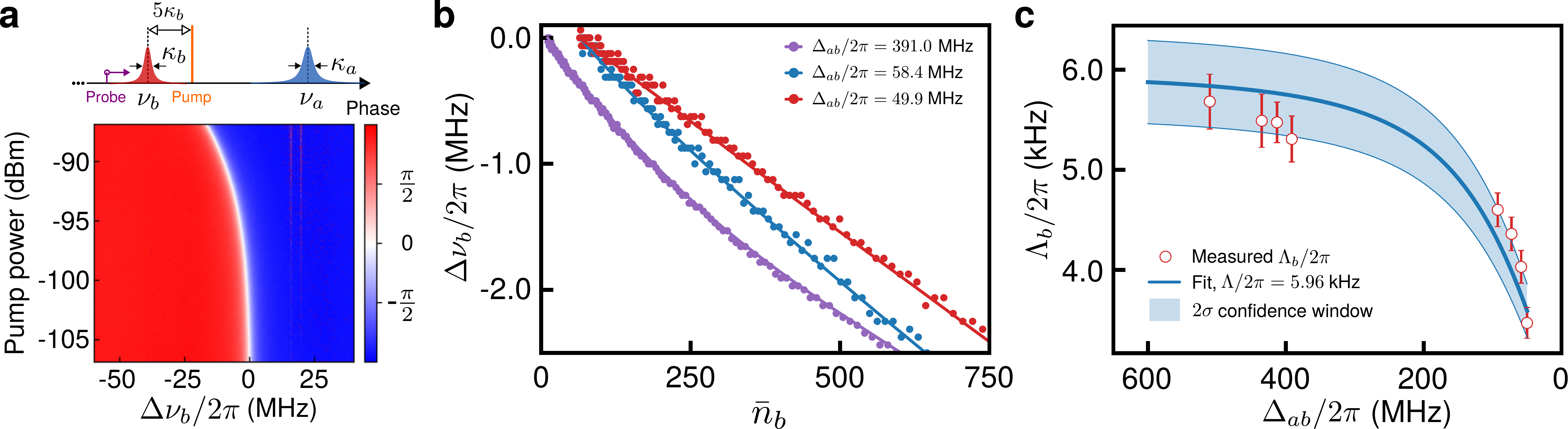}
 	\caption{\textbf{Kerr nonlinearity measurement.} (a) Left column: schematic showing the pump-probe setup. Note that modes shown are \textit{polariton} modes, not the bare modes. A strong pump tone (orange) is applied 5 linewidths positively detuned from $\hat{c}_b$ (see text), and the modified polariton resonance frequency identified using a weak probe (purple). Increasing polariton mode occupation $\bar{n}_b$ by increasing pump tone power, the frequency shift $\Delta\nu_b$ of polariton mode is measured versus $\bar{n}_b$ for various detunings between the bare linear and nonlinear modes, $\Delta_{ab} = \omega_a - \omega_b$. Lines are fits to $-\Lambda_b \bar{n}_b + \epsilon O(\bar{n}_b^2)$ where higher order terms $\propto \epsilon$ become important as pump power increases. The fits are used to extract the polariton mode nonlinearity $\Lambda_b$. With decreasing detuning $\Delta_{ab}$, the initial slope of the fit decreases, indicating a decrease in the strength of the $\hat{c}_b$ mode nonlinearity as the hybridization increases. (b) Measured Kerr nonlinearity $\Lambda_b$ of polariton mode $\hat{c}_b$ (red circles) extracted from (a), as a function of detuning $\Delta_{ab}$. Solid blue line is the best fit to Eq.~(\ref{eq:lambdab}), with the obtained fit value of $\Lambda/2\pi = 5.96$~kHz. The shaded region indicates the 2$\sigma$ confidence interval for the fit.}
 	\label{fig:kerr}
 \end{figure*}
 

We can now rewrite the Kerr nonlinear term of the full Hamiltonian, Eq.~(\ref{eq:hsys}), in the polariton basis. Writing the nonlinear term $\hat{\mathcal{H}}_{\Lambda}$ as:
\begin{align}
    \hat{\mathcal{H}}_{\Lambda} = -\frac{\Lambda}{2}\hat{b}^{\dagger}\hat{b}^{\dagger}\hat{b}\hat{b}
\end{align}
and noting from Eq.~(\ref{eq:polDef}) that:
\begin{align}
    \hat{b} = \mathbf{P}_{21} \hat{c}^{\ }_a + \mathbf{P}_{22} \hat{c}^{\ }_b = \sum_{n} \mathbf{P}_{2n}\hat{c}^{\ }_n
\end{align}
the nonlinear Hamiltonian in terms of polariton modes takes the form:
\begin{align}
    \hat{\mathcal{H}}_{\Lambda} = -\frac{\Lambda}{2}\sum_{nmrs} \mathbf{P}^*_{2n}\mathbf{P}^*_{2m}\mathbf{P}^{\ }_{2r}\mathbf{P}^{\ }_{2s} \hat{c}_n^{\dagger}\hat{c}_m^{\dagger}\hat{c}^{\ }_r\hat{c}^{\ }_s \equiv -\frac{\Lambda}{2}\sum_{nmrs} \mathcal{A}^{\ }_{nmrs} \hat{c}_n^{\dagger}\hat{c}_m^{\dagger}\hat{c}^{\ }_r\hat{c}^{\ }_s
    \label{eq:hlambda}
\end{align}
Therefore, the coupling transforms the localized nonlinearity of mode $\hat{b}$ into self- and cross-Kerr interactions between the polariton modes of the system. The Kerr-induced frequency shift observed for either polariton mode will be a combination of these terms, making it complicated to determine in general.

However, we can obtain a simplified expression by assuming operation near a stable fixed point and assuming a strong polariton mode occupation, both conditions that are expected to be valid for the typical pump-probe measurement scheme. The experimental scheme proceeds similarly to the case for a single nonlinear mode: a strong pump tone is applied to the system at a positive detuning of five linewidths away from \textit{polariton} mode $\hat{c}^{\ }_b$, predominantly pumping this mode, although also residually (weakly) pumping mode $\hat{c}_a$ (see schematic in Fig.~\ref{fig:kerr}~(a)). The resulting steady-state polariton amplitudes, and therefore occupations, can be conveniently determined by first obtaining the nonlinear and linear mode amplitudes $\bar{\beta}$, $\bar{\alpha}$ by solving Eqs.~(B6) and (B8) respectively, reproduced below:
\begin{subequations}
\begin{align}
  \left[ \left( \widetilde{\Delta}_{db} + \Lambda|\bar{\beta}|^2 \right)^2  + \frac{\widetilde{\gamma}^2}{4} \right] |\bar{\beta}|^2 &= \varg^2 |\chi_a|^2 \eta^2  \\
      \bar{\alpha} &=  - \chi_a \left( i \varg \bar{\beta} + i \eta \right)
\end{align}
\end{subequations}
recalling the renormalized nonlinear mode detuning and damping parameters respectively:
\begin{align}
    \widetilde{\Delta}_{db} &= \omega_d - (\omega_b + \varg^2 |\chi_a|^2 \Delta_{da}) \nonumber \\
    \widetilde{\gamma} &= \gamma + \gamma_{\varphi} + \varg^2 |\chi_a|^2 \kappa
\end{align}
where $\chi_a = (-i\Delta_{da} + \frac{\kappa}{2})^{-1}$. Then, the steady-state polariton amplitudes, $\bar{c}^{\ }_a$, $\bar{c}^{\ }_b$, are easily determined via the transformation matrix introduced in Eq.~(\ref{eq:polDef}):
\begin{align}
    \begin{pmatrix}
    \bar{c}^{\ }_a \\
    \bar{c}^{\ }_b 
    \end{pmatrix}
    = \mathbf{P}^{-1}
    \begin{pmatrix}
    \bar{\alpha} \\ 
    \bar{\beta}
    \end{pmatrix}
    \label{eq:polSS}
\end{align}
Finally, the application of a weak probe determines Kerr-mediated frequency shifts, as dictated by the nonlinear Hamiltonian, Eq.~(\ref{eq:hlambda}). We are only interested in shifts to the polariton mode $\hat{c}^{\ }_b$; the corresponding terms of the nonlinear Hamiltonian are given by:
\begin{align}
    \hat{\mathcal{H}}_{\Lambda} \approx -\frac{\Lambda}{2}\left[\mathcal{A}_{2222}\hat{c}^{\dagger}_b\hat{c}^{\ }_b +  4\mathcal{A}^{\ }_{2121}\hat{c}^{\dagger}_a\hat{c}^{\ }_a + 2\mathcal{A}^{\ }_{2221}\hat{c}^{\dagger}_b\hat{c}^{\ }_a + 2\mathcal{A}^{\ }_{2122}\hat{c}^{\dagger}_a\hat{c}^{\ }_b\right]\hat{c}_b^{\dagger}\hat{c}^{\ }_b + (\hat{c}^{\dagger}_a\hat{c}^{\ }_a~{\rm -only~and~non~Kerr~shift~terms})
\end{align}
We now perform a semiclassical approximation, linearizing the above Hamiltonian around the fixed point defined by Eqs.~(\ref{eq:polSS}), under which the effective Kerr-mediated shift $\Delta\nu_b$ of the polariton frequency $\nu_b$ is given by:
\begin{align}
    \Delta\nu_b = -\Lambda\left[\mathcal{A}_{2222}|\bar{c}_b|^2 +  2\mathcal{A}^{\ }_{2121}|\bar{c}_a|^2 + 2\mathcal{A}^{\ }_{2221}\bar{c}_b^*\bar{c}_a + \mathcal{A}^{\ }_{2122}\bar{c}_a^{*}\bar{c}^{\ }_b\right]
\end{align}
Finally, the effective measured Kerr constant $\Lambda_b$ is obtained by determining the frequency shift per photon occupying the polariton mode, $\bar{n}_b = |\bar{c}_b|^2$:
\begin{align}
 \Lambda_b = -\frac{\Delta\nu_b}{\bar{n}_b} = \Lambda\left[\mathcal{A}_{2222} +  2\mathcal{A}^{\ }_{2121}\frac{|\bar{c}_a|^2}{|\bar{c}_b|^2} + 2\mathcal{A}^{\ }_{2221}\frac{\bar{c}_a}{\bar{c}_b} + \mathcal{A}^{\ }_{2122}\frac{\bar{c}_a^{*}}{\bar{c}_b^*}\right]
    \label{eq:lambdab}
\end{align}
Clearly, $\Delta\nu_b$ and the measured Kerr constant $\Lambda_b$ depend on $\mathcal{A}_{nmrs}$ and consequently on the detuning between the bare linear and nonlinear modes, $\Delta_{ab} = \omega_a-\omega_b$, as well as the strength of their coupling $\varg$. As a result, both will vary as the nonlinear mode frequency $\omega_b$ is swept, even though the bare nonlinear mode Kerr constant $\Lambda$ remains unchanged. In addition to this dependence on  $\omega_b$, Eq.~(\ref{eq:lambdab}) also accounts for the small but nonzero occupation of polariton mode $\hat{c}_a$ due to this mode being weakly driven, and the corresponding cross-Kerr shifts this mediates. 

Experimentally, a single pump-probe measurement with pump frequency $\omega_P$ at a fixed nonlinear mode frequency populates the polariton mode $\hat{c}_b$ as the pump power $P$ is increased. We first calibrate the polariton mode occupation with the applied pump power via $\bar{n}_{b}=|\bar{c}_b|^2=\frac{\kappa_b}{\Delta_P^2+(\kappa_b/2)^2}\frac{P}{\hbar{\omega_P}}$, where $\kappa_b$ is the linewidth of polariton mode $\hat{c}_b$, and $\Delta_P = 5\kappa_b$ is the detuning between the pump frequency and the bare polariton mode frequency~\cite{aspelmeyer_cavity_2014}. The observed frequency shift $\Delta\nu_b$ as a function of $\bar{n}_b$ is shown in Fig.~\ref{fig:kerr}~(a) for various detunings between the bare linear and nonlinear modes $\Delta_{ab}$. By fitting the observed frequency shift to $\bar{n}_b$, we obtain the measured polariton mode Kerr constant $\Lambda_b$. Each such measurement yields $\Lambda_b$ at the given $\Delta_{ab}$. By sweeping the nonlinear mode frequency, we obtain $\Lambda_b$ as a function of $\Delta_{ab}$, with the results plotted in red in Fig.~\ref{fig:kerr}~(b). Note that as the detuning $\Delta_{ab}$ decreases, the measured Kerr nonlinearity strength also decreases, since increased hybridization dilutes the nonlinearity of the originally nonlinear mode. By fitting the experimental results to Eq.~(\ref{eq:lambdab}) with the bare nonlinearity $\Lambda$ as the only fitting parameter, we obtain the solid blue curve in Fig.~\ref{fig:kerr}~(b), with the fit value $\Lambda/2\pi = 5.96$~kHz. The shaded blue region indicates the $2\sigma$ confidence interval of the fit, which finally yields the bare nonlinearity of $\Lambda/2\pi = 5.96\pm 0.2$~kHz for Device A.

\subsection*{Typical Kerr nonlinearity strength of optical microresonators}
\label{subsec:optKerr}

In this subsection we calculate the typical Kerr nonlinearity strength, or equivalently the Kerr-mediated frequency shift per photon, for nonlinear optical microresonators. For an optical microresonator with center frequency $\omega_{\rm op}$, refractive index $n$, second-order nonlinear refractive index $n_2$, and mode volume $V_0$, the Kerr shift per photon, $\Lambda_{\rm op}$ is given by~\cite{kippenberg_dissipative_2018}:
\begin{align}
    \Lambda_{\rm op} = \frac{\hbar\omega_0^2 c n_2}{n^2 V_0}
\end{align}
where $c$ is the speed of light in vacuum. Using parameter values for silicon nitride optical microresonators~\cite{gaeta_photonic-chip-based_2019} - a popular and successful material choice - we have: $\omega_{\rm op}/(2\pi) = 100~{\rm THz}$ (equivalently, wavelength $\lambda \simeq 1.55~\mu{\rm m}$), $n = 2$, $n_2 = 2.5 \times 10^{-19}~{\rm m^2~W^{-1}}$, and $V_0 = (\lambda/n)^3$, we obtain:
\begin{align}
    \Lambda_{\rm o}/(2\pi) \simeq 100~{\rm Hz}
\end{align}
which is about two orders of magnitude lower than the realized $\Lambda$ for Device A. Optical microresonators are engineered to have high quality factors; we consider a large value of $Q \simeq 10^7$. For $\omega_{\rm op}/(2\pi) = 100~{\rm THz}$, this implies microresonator loss rates of $\kappa_{\rm op}/(2\pi) \simeq 10~{\rm MHz}$. As a result, the ratio of $\Lambda_{\rm op}$ to the loss rate is $\Lambda_{\rm op}/\kappa_{\rm op} \simeq 10^{-5}$, again about two orders of magnitude smaller than the smallest value realized by devices in our experiment.

\section*{Simulating stochastic differential equations}
\label{sec:sdes}

In this section, we provide details of the numerical simulations of the SDEs (Eqs.~(3) of the manuscript), and in particular how these are used to calculate quantities of interest such as the temporal coherence functions. For convenience, we reproduce the set of SDEs below:
\begin{align}
    d\vec{\zeta} = \vec{A}_{\rm c}(\vec{\zeta})dt + \mathbf{B}_{\rm st}(\vec{\zeta},\Lambda,\gamma_{\varphi})d\vec{W}(t)
    \label{eq:sdes}
\end{align}
where $\mathbf{B}_{\rm st}$ is the matrix square root of the diffusion matrix, defined via $\mathbf{D}_{\rm st} = \mathbf{B}_{\rm st} \mathbf{B}_{\rm st}^T$. For a 4-by-4 diffusion matrix $\mathbf{D}$, the noise matrix $\mathbf{B}$ is not unique; it is in general a 4-by-$k$ non-square matrix, with $d\vec{W}(t)$ then being a $k$-by-1 vector of independent Wiener increments. While this freedom of choice in the noise matrix can be used to improve SDE convergence properties~\cite{drummond_quantum_2003}, we find that here a square matrix ($k=4$) suffices. We write it in the form:
\begin{align}
    \mathbf{B}_{\rm st} = \sqrt{\Gamma}~\mathbf{B}_1 + \sqrt{\gamma_{\varphi}}~\mathbf{B}_2 = 
    \sqrt{\Gamma}
    \begin{pmatrix}
    \mathbf{0} & \mathbf{0}  \\
    \mathbf{b}_1 & \mathbf{0} \\
    \end{pmatrix}
    +
    \sqrt{\gamma_{\varphi}}
    \begin{pmatrix}
    \mathbf{0} & \mathbf{0}  \\
    \mathbf{0} & \mathbf{b}_2 \\
    \end{pmatrix}
\end{align}
where $\mathbf{0}$ is the 2-by-2 matrix of zeros as before, and $\mathbf{B}_1$, $\mathbf{B}_2$ are the noise matrices introduced in the main text. Here we also provide their explicit forms in terms of the 2-by-2 component matrices $\mathbf{b}_1$ and $\mathbf{b}_2$:
\begin{align}
    \mathbf{b}_1 = 
    \begin{pmatrix}
    e^{i\theta/2}\beta & 0 \\
    0 & e^{-i\theta/2}\beta^{\dagger}
    \end{pmatrix},~
    \mathbf{b}_2 = 
    \sqrt{\frac{\beta^{\dagger}\beta}{2}}
    \begin{pmatrix}
    e^{i\pi/4} & e^{-i\pi/4} \\
    e^{-i\pi/4} & e^{i\pi/4}
    \end{pmatrix}
\end{align}
Finally, we have defined the parameters $\Gamma$ and $\theta$ via:
\begin{align}
    \Gamma e^{i\theta} \equiv i\Lambda -\gamma_{\varphi} \implies \Gamma = \sqrt{\Lambda^2 + \gamma_{\varphi}^2},~\theta = \arctan\left( -\frac{\Lambda}{\gamma_{\varphi}} \right).
\end{align}
The validity of the noise matrix $\mathbf{B}_{\rm st}$ as the square root of the diffusion matrix $\mathbf{D}_{\rm st}$ may be easily verified by direct multiplication. 

\subsection*{Practical computation of steady-state operator moments and correlation functions using SDEs}

Simulations of the SDEs in Eq.~(\ref{eq:sdes}) yield individual stochastic trajectories of the stochastic variables $\vec{\zeta}(t)$, which may then be used to compile \textit{normal-ordered} moments and correlation functions. In what follows, we use expressions for moments and correlation functions for the linear mode as examples, since these are directly accessible via experiment. However the expressions hold equally for nonlinear mode operators by appropriate substitutions.

Suppose Eqs.~(\ref{eq:sdes}) are solved to obtain $N_s$ stochastic trajectories, yielding a set of stochastic trajectories $\{\vec{\zeta}_i(t)\}$ for $i = 1,\ldots, N_s$. Then, first-order moments for the linear mode may be determined via stochastic averaging (indicated by notation $\avg{\cdot}_s$) as follows:
\begin{align}
    \avg{\hat{a}(t)} &= \langle \alpha(t) \rangle_s = \lim_{N_s \to \infty} \frac{1}{N_s}\sum_{i=1}^{N_s} \alpha_i(t) 
\end{align}
Normal-ordered two-time correlation functions follow similarly:
\begin{subequations}
\begin{align}
    \avg{\hat{a}^{\dagger}(t+\tau)\hat{a}(t)} &= \langle \alpha^{\dagger}(t+\tau)\alpha(t) \rangle_s = \lim_{N_s \to \infty} \frac{1}{N_s}\sum_{i=1}^{N_s} \alpha_i^{\dagger}(t+\tau)\alpha_i(t) \\
    \avg{\hat{a}(t+\tau)\hat{a}(t)} &= \langle \alpha(t+\tau)\alpha(t) \rangle_s = \lim_{N_s \to \infty} \frac{1}{N_s}\sum_{i=1}^{N_s} \alpha_i(t+\tau)\alpha_i(t)
\end{align}
\end{subequations}
Note that the other normal-ordered, time anti-ordered correlation functions $\avg{\hat{a}^{\dagger}(t)\hat{a}(t+\tau)}$ and $\avg{\hat{a}^{\dagger}(t)\hat{a}^{\dagger}(t+\tau)}$ may be obtained from the two expressions above respectively by conjugation.

While the above expressions allow access to moments and correlation functions at arbitrary times, when analyzing the long time coherence of the emergent frequency combs we will ultimately be interested in steady state quantities. The requirement of a steady state allows an alternative calculation of the above quantities. To directly acquire steady state quantities, we simulate Eqs.~(\ref{eq:sdes}) for times $t \in [0, t_{\rm ss}+T_W]$ with simulation time step $\Delta t$, and retain solutions in the time window $t \in [t_{\rm ss},t_{\rm ss}+T_W]$ where $T_W$ is the length of time window. The time $t_{\rm ss}$ is chosen long enough that the solutions $\{\vec{\zeta}_i(t)\}$ within the stored window are extracted when initial transients have decayed away; this initial $t_{\rm ss}$ value is verified self-consistently, as discussed at the end of this section. For simplicity, we now index the solutions in this time window by times $t_j \in [t_{\rm ss}, t_{\rm ss} + T_W]$, such that $t_j = t_{\rm ss} + j \Delta t$ for $j = 0,\ldots,M_1$, where $M_1 = T_W/\Delta t$. 

Since in the steady state first order moments should become stationary in time, we can equivalently average moments over time, with the results being equivalent to ensemble averaging if a true steady state has been achieved. In practice, to take advantage of parallelization available with modern computing clusters, we compute moments by averaging over \textit{both} trajectories and time:
\begin{align}
    \avg{\hat{a}} \approx \frac{1}{N_s}\sum_{i=1}^{N_s} \left[ \frac{1}{M_1+1}\sum_{j=0}^{M_1} \alpha_i(t_j) \right]
\end{align}
where the term in square bracket implements time-averaging.

Similarly, two-time correlation functions reduce to single time quantities in the steady state. Suppose we wish to compute correlation functions such as $\avg{\hat{a}^{\dagger}(\tau)\hat{a}(0)}$ for $\tau \in [0, T_A]$ where $T_A \leq T_W$. The time average in this case is now performed over a subset of the total window of length $T$, namely over $t_j$ where $j = 0,\ldots,M_2$ where $M_2 = (T_W-T_A)/\Delta t \leq M_1$. Then, steady state correlation functions may be obtained by ensemble and time averaging via:
\begin{align}
    \avg{\hat{a}^{\dagger}(\tau)\hat{a}(0)} \approx \frac{1}{N_s}\sum_{i=1}^{N_s} \left[ \frac{1}{M_2+1 }\sum_{j=0}^{M_2} \alpha_i^{\dagger}(t_j+\tau)\alpha_i(t_j) \right]
\end{align}
Note that if $T_A = T_W$, the required correlation function spans the entire retained window of length $T_W$; only a single correlation function is obtained $(M_2 = 0)$ and thus time averaging has no effect. In practice, we retain solutions for a time window $T_W$ that is larger than the required length of the correlation function $T_A$, so that time-averaging can be performed.

Finally, to verify that all averaged results computed above are truly steady state quantities, we increase the value of $t_{\rm ss}$ beyond its chosen initial value and recompute the results, checking to see whether the averaged quantities are unchanged. If so, they are independent of $t_{\rm ss}$ and we can be confident of having computed steady state quantities. Otherwise, the procedure is repeated for increasing $t_{\rm ss}$ values until this condition is met. 

\subsection*{Calculating the filtered output temporal coherence function}
\label{sec:calcCorr}

For calculations of comb coherence, we introduce in the main text the first-order temporal coherence function $G^{(1)}(\tau)$; here we rewrite it in a slightly different but ultimately equivalent form:
\begin{align}
    G^{(1)}(\tau) = \frac{\avg{\Delta\hat{I}(0)\Delta\hat{I}(\tau)}}{\avg{\Delta\hat{I}(0)\Delta\hat{I}(0)}} = \frac{\avg{\hat{I}(0)\hat{I}(\tau)} - \avg{\hat{I}(0)}^2}{ \avg{\hat{I}(0)\hat{I}(0)} - \avg{\hat{I}(0)}^2 }
    \label{eq:G1}
\end{align}
where $\hat{I}(t)$ is the \textit{measured} cavity output quadrature, and we have introduced \textit{reduced} steady-state correlation functions for arbitrary operators $\hat{o}_1, \hat{o}_2$ as:
\begin{align}
    \avg{\Delta\hat{o}_1(0)\Delta\hat{o}_2(\tau)} = \avg{ (\hat{o}_1(0)-\avg{\hat{o}}_1)(\hat{o}_2(\tau)-\avg{\hat{o}}_2) } = \avg{\hat{o}_1(0)\hat{o}_2(\tau)} - \avg{\hat{o}_1}\avg{\hat{o}_2}
\end{align}

Note, however, that simulating the SDEs of Eqs.~(\ref{eq:sdes}) only yields \textit{intra}cavity quantities, while we require measured cavity \textit{output} quantities to compute $G^{(1)}(\tau)$. In this section, we show how the two can be related using quantum input-output theory~\cite{gardiner_input_1985}. We begin by analyzing the steady state output quadrature correlation function:
\begin{align}
    \avg{\hat{i}(0)\hat{i}(\tau)} = \avg{\hat{i}(\tau)\hat{i}(0)}^*
\end{align}
where $\hat{i}(t)$ is the cavity mode output quadrature \textit{prior to any post-processing} (in particular downconversion and demodulation) carried out in the experiment. It can be written in terms of the output field non-Hermitian operators $\hat{a}_{\rm out}(t)$:
\begin{align}
    \hat{i}(t) = \frac{1}{\sqrt{2}}\left( \hat{a}_{\rm out}(t) + \hat{a}_{\rm out}^{\dagger}(t) \right)
\end{align}
In terms of the non-Hermitian output operators, the output quadrature correlation function takes the form:
\begin{align}
    \avg{\hat{i}(0)\hat{i}(\tau)} = \frac{1}{2}\left( \avg{\hat{a}_{\rm out}(0)\hat{a}_{\rm out}(\tau)} + \avg{\hat{a}^{\dagger}_{\rm out}(0)\hat{a}^{\dagger}_{\rm out}(\tau)} + \avg{\hat{a}^{\dagger}_{\rm out}(0)\hat{a}_{\rm out}(\tau)} + \avg{\hat{a}_{\rm out}(0)\hat{a}_{\rm out}^{\dagger}(\tau)} \right)
\end{align}
It now proves useful to normal-order and time anti-order the individual correlation functions. This requires use of the commutation relationships between the non-Hermitian output operators~\cite{gardiner_input_1985}:
\begin{align}
    [\hat{a}_{\rm out}(t),\hat{a}_{\rm out}(t')] = 0 = [\hat{a}^{\dagger}_{\rm out}(t),\hat{a}^{\dagger}_{\rm out}(t')],~[\hat{a}_{\rm out}(t),\hat{a}_{\rm out}(t')] = \delta(t-t')
\end{align}
Then, the output quadrature correlation function becomes:
\begin{align}
    \avg{\hat{i}(0)\hat{i}(\tau)} = \frac{1}{2}\left( \avg{\hat{a}_{\rm out}(\tau)\hat{a}_{\rm out}(0)} + \avg{\hat{a}^{\dagger}_{\rm out}(0)\hat{a}^{\dagger}_{\rm out}(\tau)} + \avg{\hat{a}^{\dagger}_{\rm out}(0)\hat{a}_{\rm out}(\tau)} + \avg{\hat{a}_{\rm out}^{\dagger}(\tau)\hat{a}_{\rm out}(0)} + \delta(\tau) \right)
\end{align}
Note that the second and fourth terms in the expressions are simply conjugates of the first and third terms respectively. 

To calculate the reduced correlation function, we require the steady state quantity:
\begin{align}
    \avg{\hat{i}(0)}\avg{\hat{i}(\tau)} = \frac{1}{2}\left( \avg{\hat{a}_{\rm out}(0)}\avg{\hat{a}_{\rm out}(\tau)} + \avg{\hat{a}^{\dagger}_{\rm out}(0)}\avg{\hat{a}^{\dagger}_{\rm out}(\tau)} + \avg{\hat{a}^{\dagger}_{\rm out}(0)}\avg{\hat{a}_{\rm out}(\tau)} + \avg{\hat{a}_{\rm out}^{\dagger}(\tau)}\avg{\hat{a}_{\rm out}(0)} \right)
\end{align}
Using the above, we can finally write the reduced output quadrature correlation function as:
\begin{align}
    \avg{\Delta\hat{i}(0)\Delta\hat{i}(\tau)} = \frac{1}{2}\left( \avg{\Delta\hat{a}_{\rm out}(\tau)\Delta\hat{a}_{\rm out}(0)} + \avg{\Delta\hat{a}^{\dagger}_{\rm out}(0)\Delta\hat{a}^{\dagger}_{\rm out}(\tau)} + \avg{\Delta\hat{a}^{\dagger}_{\rm out}(0)\Delta\hat{a}_{\rm out}(\tau)} + \avg{\Delta\hat{a}_{\rm out}^{\dagger}(\tau)\Delta\hat{a}_{\rm out}(0)} + \delta(\tau) \right)
\end{align}

Now we can relate the output field operators to intracavity operators via input-output theory:
\begin{align}
    \hat{a}_{\rm out}(t) = \hat{a}_{\rm in}(t) + \sqrt{\kappa}\hat{a}(t)
\end{align}
The two independent reduced output field correlation functions above can be related to the intracavity field reduced correlation functions (assuming zero temperature):
\begin{subequations}
\begin{align}
    \avg{\Delta\hat{a}_{\rm out}(\tau)\Delta\hat{a}_{\rm out}(0)} &= \kappa \avg{\Delta\hat{a}(\tau)\Delta\hat{a}(0)}  \\
    \avg{\Delta\hat{a}^{\dagger}_{\rm out}(\tau)\Delta\hat{a}_{\rm out}(0)} &= \kappa \avg{\Delta\hat{a}^{\dagger}(\tau)\Delta\hat{a}(0)}
\end{align}
\end{subequations}
Finally, the reduced \textit{output quadrature} correlation function can be related to \textit{normal-ordered intracavity} correlation functions as:
\begin{align}
    \avg{\Delta\hat{i}(0)\Delta\hat{i}(\tau)} = \frac{1}{2}\delta(\tau) + \frac{\kappa}{2}\Big( \avg{\Delta\hat{a}(\tau)\Delta\hat{a}(0)}  + \avg{\Delta\hat{a}^{\dagger}(\tau)\Delta\hat{a}(0)} + c.c. \Big)
    \label{eq:cavCorr}
\end{align}
Recall that we are ultimately interested in the correlation function for the \textit{measured} cavity output quadrature $\hat{I}(t)$, as defined in the main text, which is related to $\hat{i}(t)$ by a downconversion and demodulation step. Fortunately, it is possible to relate measured correlation functions post-filtering directly to output correlation functions prior to filtering~\cite{da_silva_schemes_2010}:
\begin{align}
    \avg{\Delta\hat{I}(0)\Delta\hat{I}(\tau)} = \mathcal{F}(\tau) \ast \avg{\Delta\hat{i}(0)\Delta\hat{i}(\tau)}
    \label{eq:filtCorr}
\end{align}
where $\mathcal{F}(\tau)$ is the composite filter function describing both downconversion and demodulation of the cavity output in the process of measurement, and $\ast$ indicates the convolution operation. Therefore, to calculate the measured $G^{(1)}(\tau)$ numerically, we first simulate Eqs.~(\ref{eq:sdes}) and calculate the reduced intracavity correlation functions on the right hand side of Eq.~(\ref{eq:cavCorr}), as described in the previous section. This enables us to obtain the output correlation function $\avg{\Delta\hat{i}(0)\Delta\hat{i}(\tau)}$. The resulting function is then passed through (i.e. convolved with) the composite filter $\mathcal{F}(\tau)$ to obtain the filtered correlation function, Eq.~(\ref{eq:filtCorr}). Finally, employing Eq.~(\ref{eq:G1}) yields the required temporal coherence function numerically.

Accounting for the filtering process is important to obtain agreement between the calculated and measured coherence functions, in particular the oscillation frequency which would otherwise differ from the experiment by $\sim 100~$MHz, the downconversion offset implemented as part of the post-processing. This is particularly evident in comparisons of the measured and numerically calculated $G^{(1)}(\tau)$ shown in Fig. 3(c) of the main text, We also note that the filtering process replaces the somewhat unphysical $\delta$-function contribution in Eq.~(\ref{eq:cavCorr}) - arising from the abstract construct of white noise in the cavity output field - with a finite quantity, as is expected for any real detection scheme which possesses a finite bandwidth. 

\subsection*{Non-exponential signatures in phase decoherence}
\label{sec:nonexp}

\begin{figure*}[t]
 	\includegraphics[scale = 0.3]{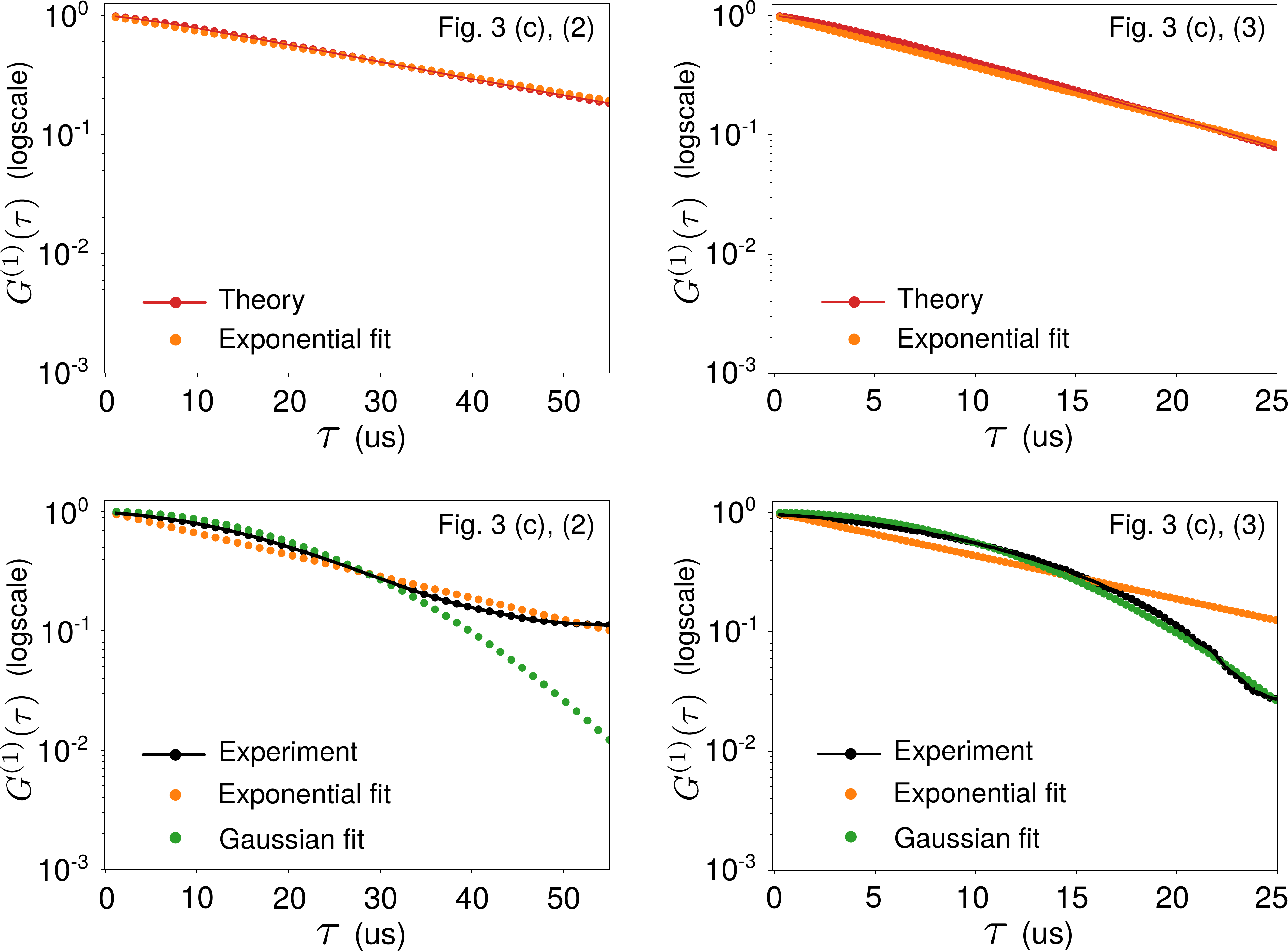}
 	\caption{\textbf{Non-exponential signatures of phase decoherence.} Top panels: Theoretically calculated decay of $G^{(1)}(\tau)$ peaks as a function of time in logscale (red), and exponential fit (orange). Left and right plots correspond to $G^{(1)}(\tau)$ plotted in Fig.~3(c), positions (2) and (3) respectively of the main text. Lower panels: Same as the top panel, but now showing experimentally obtained decay of $G^{(1)}(\tau)$ peaks as a function of time (black). In addition to the exponential fit (orange), a Gaussian fit is also shown (green), corresponding to contributions from $1/f$ noise. }
 	\label{fig:nonexp}
 \end{figure*}

Generally, the multiplicative nature of the noise described by Eqs.~(\ref{eq:sdes}) allows for nonexponential decay of the phase coherence captured by $G^{(1)}(\tau)$. Signatures of this can be better seen by extracting the theoretically calculated coherence function peaks and plotting in logscale, as shown in Fig.~\ref{fig:nonexp}, top panel. However, we find numerically that these nonexponential signatures are minor in the explored parameter regime, as can be seen by the very small deviation from a straight line in logscale. As a result the theoretical decay envelope can be considered to be exponential to a very good approximation.


For the experimentally-obtained coherence function, we have also observed that the decay envelope of the correlation function is not always perfectly exponential. We believe that the observed experimental decay is further complicated by signatures of $1/f$ noise in the system, which would lead to a Gaussian decay envelope. From typical plots of the coherence function peaks as a function of time in logscale, we find that the actual experimental decay envelope is quite close to exponential, but with some signatures of Gaussian decay. Fig. 3(c) at position 3 is a somewhat atypical example, with more pronounced nonexponential features; we usually find decay characteristics more similar to Fig. 3(c), position 2. 

Various techniques have been developed to characterize environmental noise in nature and artificial spin systems [Nature Physics 7, 565(2011), Nature Communications 4, 2337 (2013),  Nature Physics, 15, 1123 (2019)]. However, the efficacy of these methods requires sensitivity to the specific frequency distribution of the noise, e.g.: large anharmonicity, thus limiting our ability to remove this noise contribution from the experimental results, or include its effect on the theoretical calculations. Noting further that we generally find the experimental decay to be close to exponential just like the theoretical results in this parameter regime, we have used an exponentially decaying envelope to fit to both theory and experiment results and extract the coherence time. Furthermore, adding a probing system to characterize the environmental noise can also be a future research about quantum coherence in the unstable regime.

\section*{Dephasing in the weak-driving regime}
\label{sec:deph}

In this work we include the effects of flux noise on the tunable nonlinear mode via the pure dephasing term $\propto \gamma_{\varphi}$ in the system master equation. In the main text, the impact of pure dephasing on frequency comb coherence was assessed. In this appendix section we consider the influence of pure dephasing in the regime of \textit{weak} driving, far from the instability regions where frequency combs emerge. The qualitative features of this regime can be seen by neglecting the nonlinearity, which then enables an exact analysis of the dynamics. However, being able to access the dynamics in this weak driving regime is not straightforward, as we discuss in the following sections.

\subsection*{Exact dynamics }

In this linear regime, we find that the full quantum two-mode model can be reduced to a closed set of linear equations for the first and second order moments of the two modes. In particular, the linear system becomes:
\begin{align}
    \frac{d}{dt}\vec{v} = \mathbf{M} \vec{v} + \vec{d}
    \label{eq:dephTD}
\end{align}
where $\vec{v}$ is the vector of first and second order moments, and $\vec{d}$ describes the drive on the linear mode:
\begin{align}
    \vec{v} = 
    \begin{pmatrix}
    \avg{\hat{a}} \\
    \avg{\hat{a}^{\dagger}} \\
    \avg{\hat{b}} \\
    \avg{\hat{b}^{\dagger}} \\
    \avg{\hat{a}^{\dagger}\hat{a}} \\
    \avg{\hat{b}^{\dagger}\hat{b}} \\
    \avg{\hat{a}^{\dagger}\hat{b}} \\
    \avg{\hat{b}^{\dagger}\hat{a}} 
    \end{pmatrix},~
    \vec{d} = 
    \begin{pmatrix}
    -i \eta \\
    +i\eta \\
    0 \\
    0 \\
    0 \\
    0 \\
    0 \\
    0
    \end{pmatrix}
\end{align}
The dynamical matrix $\mathbf{M}$ takes the block form:
\begin{align}
    \mathbf{M} = 
    \begin{bmatrix}
    \mathbf{M}_1 & \mathbf{0} \\
    \mathbf{N} & \mathbf{M}_2
    \end{bmatrix}
\end{align}
where:
\begin{subequations}
\begin{align}
    \mathbf{M}_1 &= 
    \begin{pmatrix}
    i\Delta_{da} - \frac{\kappa}{2} & 0 & -i\varg & 0 \\
    0 & -i\Delta_{da} - \frac{\kappa}{2} & 0 & i\varg \\
    -i\varg & 0 & i\Delta_{db} - \frac{1}{2}\left(\gamma+\gamma_{\varphi}\right) & 0 \\
    0 & i\varg &  0 & -i\Delta_{db} - \frac{1}{2}\left(\gamma+\gamma_{\varphi}\right)  \\
    \end{pmatrix} \\
    \mathbf{M}_2 &= 
    \begin{pmatrix}
    -\kappa & 0 & i\varg & -i\varg \\
    0 & -\gamma & -i\varg & i\varg \\
    i\varg & -i\varg & i\Delta_{da}-i\Delta_{db} - \frac{1}{2}\left(\kappa+\gamma+\gamma_{\varphi}\right) & 0 \\
    -i\varg & i\varg &  0 & i\Delta_{db}-i\Delta_{da} - \frac{1}{2}\left(\kappa+\gamma+\gamma_{\varphi}\right) \\
    \end{pmatrix} \\
    \mathbf{N} &=
    \begin{pmatrix}
    i\eta & -i\eta & 0 & 0 \\
    0 & 0 & 0 & 0 \\
    0 & 0 & 0 & -i\eta \\
    0 & 0 & i\eta & 0
    \end{pmatrix}
\end{align}
\end{subequations}
For convenience, we define the $\hat{I}_j$ and $\hat{Q}_j$ quadratures for mode $j$ as:
\begin{align}
    \hat{I}_j = \frac{1}{\sqrt{2}}\left( \hat{d}_j + \hat{d}_j^{\dagger} \right),~\hat{Q}_j = \frac{-i}{\sqrt{2}}\left( \hat{d}_j - \hat{d}_j^{\dagger} \right) 
\end{align}
where $\hat{d}_j \in \{\hat{a},\hat{b}\}$ for $j=a,b$ respectively. The above definitions also imply that:
\begin{align}
    \hat{A}_j^2 \equiv \hat{I}_j^2 + \hat{Q}_j^2 = \avg{\hat{d}_j^{\dagger}\hat{d}_j}
    \label{eq:A}
\end{align}

Removing the coupling $(\varg=0)$ renders the undriven $(\eta = 0)$ dynamical matrix diagonal, and the system decay rates can simply be read off. The linear mode decay rate for $\avg{\hat{a}}$ is $\frac{\kappa}{2}$ and for $\avg{\hat{a}^{\dagger}\hat{a}}$ is $\kappa$, indicating that the cavity mode experiences no pure dephasing. In contrast, the nonlinear mode amplitude $\avg{\hat{b}}$ decays at the rate $(\gamma+\gamma_{\varphi})/2$, while the nonlinear mode occupation decays at the rate $\gamma$. Therefore in this linear dynamical regime it should in principle be possible to determine the dephasing rate $\gamma_{\varphi}$ by observing the decay of the nonlinear mode quadrature $\hat{I}_b$, in comparison to the decay of $\avg{\hat{I}_b^2+\hat{Q}_b^2}$.

\subsection*{Dephasing measurement in the two-level approximation}

When the anharmonicity $\Lambda$ of the nonlinear mode is large compared to its damping rate $\gamma$, the nonlinear mode may be accurately modeled as a two-level system. In this regime, a standard approach to measuring the two relevant decay rates has been readily employed in cQED: by tuning the resulting two-level system frequency far from the cavity mode, a dispersive coupling between the two-level system and the cavity mode is realized, which enables mapping the two-level system state to one of two cavity pointer states. Then, a measurement of the two-level system state is made via a homodyne measurement of the cavity output field. The effects of dephasing in this regime can then be recast into a form familiar in cavity QED: it leads to an additional depolarization of $\avg{\hat{\sigma}_x}$ (analogous to $\avg{\hat{I}_b}$), without affecting the relaxation rate of $\avg{\hat{\sigma}_z}$ (analogous to $\avg{\hat{I}_b^2 + \hat{Q}_b^2})$. A standard Ramsey experiment yields the depolarization rate $(\gamma+\gamma_{\varphi})/2 = (T_2^*)^{-1}$ for $\avg{\hat{\sigma}_x}$, and by obtaining the relaxation rate $\gamma = T_1^{-1}$ for $\avg{\hat{\sigma}_z}$, one can extract the pure dephasing rate $\gamma_{\varphi}/2 \equiv (2T_{\varphi})^{-1} =  (T_2^*)^{-1} - (2T_1)^{-1}$.

By mapping the two-level system state to cavity pointer states, measurements in the dispersive regime enable access to two-level system dynamics that occur on longer timescales set by $1/\gamma$, even though measurements are being made of the cavity which evolves on a much shorter time scale $1/\kappa$, since $\kappa \gg \gamma$. The latter condition is in fact necessary to ensure a measurement time that is shorter than the relaxation time of the nonlinear mode, which reduces errors due to unwanted relaxation between the end of any evolution of interest of the two-level system and the conclusion of the measurement of its state. Furthermore, since this approach measures the time-\textit{integrated} homodyne current to obtain the two-level system dynamics, its temporal resolution is not limited by the DAC (digital-to-analog convertor) that determines the temporal resolution of the obtained homodyne voltage; instead, the temporal resolution is set by the degree of control over microwave pulse generation for the manipulation of the two-level system and the cavity mode.

However, for the devices under study here, we are precisely interested in the weakly-nonlinear regime, where $\Lambda/\gamma \ll 1$. While necessary for the observation of coherent frequency combs, this renders addressing just two states of the nonlinear mode unfeasible, and thus rules out making measurements of the nonlinear mode via a dispersive coupling to the cavity mode. In this case, one must resort to a direct temporal measurement of the relaxation of moments, which we discuss in the next section.

\subsection*{Cavity ringdown method and theoretical simulations}

Even when a two-level description of the nonlinear mode is not feasible, Eqs.~(\ref{eq:dephTD}) indicate that under weak driving it should still be possible to observe the effect of pure dephasing on the nonlinear mode moments. To do so, one would ideally like to probe the nonlinear mode dynamics directly, without having to observe the linear mode. This requires effectively decoupling the nonlinear mode from the linear mode (by being detuned far away) while still retaining a coupling to the outside world. However, the 3-D transmon design isolates the nonlinear mode from a direct coupling to the environment, successfully allowing for a much higher-$Q$ nonlinear mode than lumped-element or coplanar waveguide architectures. While this design usefully reduces both the relaxation rate $\gamma$ and pure dephasing rate $\gamma_{\varphi}$~\cite{paik_transmon_2011}, it also means that we only have direct access to linear mode quadratures, $\hat{I}_a$, $\hat{Q}_a$.

 As such, one is restricted to determining the dephasing rate by monitoring moments of cavity quadratures. The approach one would employ is a ringdown setup~\cite{maillet_classical_2016}: a coherent drive is placed on the system to initialize it to a nontrivial state in phase space, following which the drive is turned off and the resulting ringdown dynamics of the measured first and second order cavity moments recorded as the two-mode system returns to the undriven steady-state. Comparing the rates of relaxation for first and second order moments then enables a calculation of the pure dephasing rate $\gamma_{\varphi}$.


\begin{figure*}[t]
 	\includegraphics[scale = 0.39]{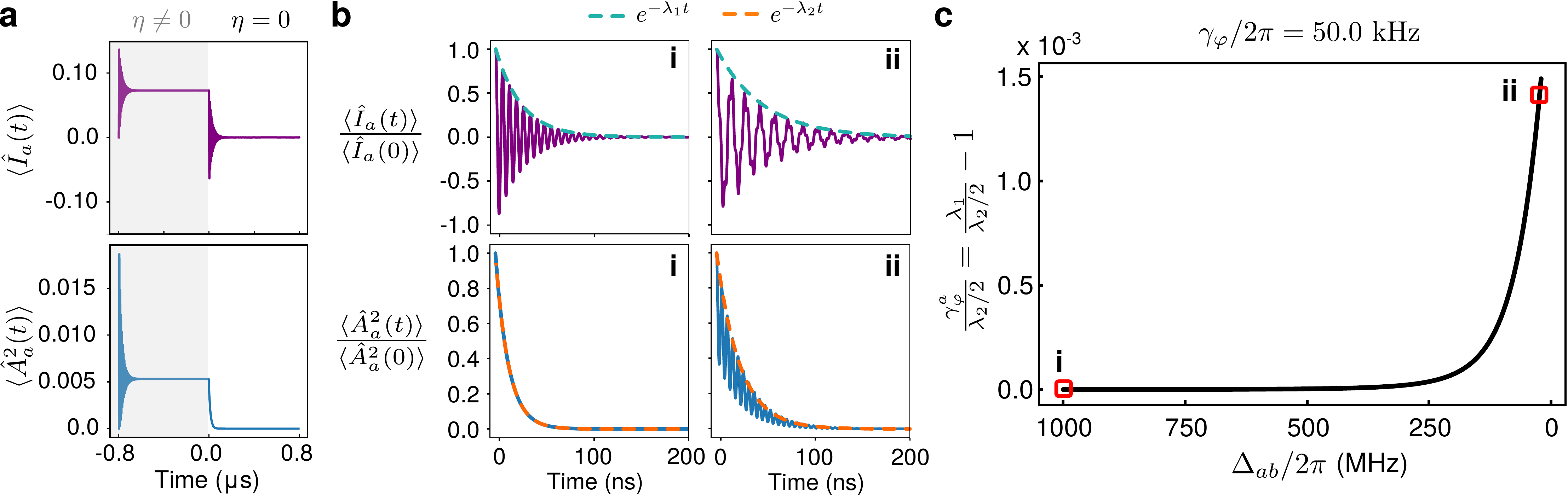}
 	\caption{\textbf{Theoretical analysis of dephasing in the linear regime via cavity ringdown.} (a) Typical dynamics of cavity quadrature moments under the initialization drive (gray region) and when the drive is turned off (blank region), displaying ringdown. (b) (i) and (ii) show ringdown dynamics for $\avg{\hat{I}_a(t)}$ (top panel) and $\avg{\hat{A}_a(t)}$ (bottom panel) at the indicated values of detuning $\Delta_{ab}$ between the linear and nonlinear mode in (c). Also shown are fits to exponential decays; the top (bottom) panel shows fits with rates determined by the eigenvalue $\lambda_1$ ($\lambda_2$) for $\avg{\hat{I}_a(t)}$ ($\avg{\hat{A}_a(t)}$), computed from the dynamical matrix $\mathbf{M}_1$ ($\mathbf{M}_2$) in the ringdown regime ($\eta=0$). (c) Extracted decay rates from ringdown dynamics as a function of $\Delta_{ab}$. For (b), (c), we choose the actual nonlinear mode dephasing rate $\gamma_{\varphi}/2\pi = 50.0~$kHz, which is larger than the estimated dephasing rates in the comb generation regimes for both experimental devices.}
 	\label{fig:deph}
 \end{figure*}


 To explore the feasibility of such an approach, we perform numerical simulations of Eqs.~(\ref{eq:dephTD}) under this ringdown setup. We assume a much larger pure dephasing rate $\gamma_{\varphi}/2\pi = 50.0~{\rm kHz}$ than estimated for either of our devices, for reasons that will become clear shortly. The typical initialization and ringdown evolution is shown in Fig.~\ref{fig:deph}~(a). The drive $\eta$ is turned on at $t = -0.8~\mu$s, and then turned off at $t=0$, following which the cavity undergoes relaxation to return to the undriven steady-state. The ringdown dynamics are shown in Fig.~\ref{fig:deph}~(b) for two different detunings between the linear and nonlinear modes, $\Delta_{ab}$. We fit exponentials with decay constants $\lambda_1$, $\lambda_2$ to the moments $\avg{\hat{I}_a(t)}$, $\avg{\hat{A}_a^2(t)}$ (see Eq.~(\ref{eq:A}) respectively, and define the \textit{dephasing rate experienced by the linear mode} as $\gamma_{\varphi}^a = \lambda_1 - \lambda_2/2$. When the detuning is large compared to the coupling (i), the linear and nonlinear modes are effectively decoupled, so that the linear mode should experience no pure dephasing and $\gamma_{\varphi}^a$ is vanishingly small. With decreasing detuning (ii), the linear and nonlinear modes hybridizes, and the linear mode inherits some dephasing, so that $\gamma_{\varphi}^a$ increases. 
 
 The dephasing experienced by the linear mode $\gamma_{\varphi}^a$ as a function of $\Delta_{ab}$ is plotted in Fig.~\ref{fig:deph}~(c), scaled by $\lambda_2/2$. While in principle such an approach may be used to extract the pure dephasing rate $\gamma_{\varphi}$, Fig.~\ref{fig:deph}~(c) brings to light a number of technical difficulties. Firstly, the variation due to pure dephasing is superimposed on the very fast cavity decay rate; the \textit{relative} difference in decay rates $\sim \gamma_{\varphi}/\kappa$ is therefore very small and difficult to extract experimentally, even though we have assumed a dephasing rate here much larger than those obtained in the main text. In contrast, spectroscopy of the two-level system compares $\gamma_{\varphi}$ directly to $\gamma$. Secondly, since this is a direct temporal measurement, its accuracy is limited by the DAC resolution. Small changes in the very short cavity relaxation time are therefore more uncertain.

Both these issues mean that obtaining the pure dephasing rate from direct cavity ringdown measurements under moderate to strong hybridization is likely to be inaccurate. As a result, we instead employ the strategy of obtaining $\gamma_{\varphi}$ in the \textit{nonlinear} regime, in particular within the frequency comb regime. Here, the effect of the bare mode decay rates $\gamma$ and $\kappa$ is overcome since the system starts to undergo self-oscillation, as discussed in the main text. Then, the comb coherence is limited entirely by the nonlinearity strength and the pure dephasing rate. By measuring the nonlinearity strength via a pump-probe measurement of the hybridized system, as discussed in Section~\ref{ssec:kerr}, we are able to use SDE simulations of comb coherence to obtain an estimate of $\gamma_{\varphi}$.

\section*{Derivation of Floquet eigensystem}
\label{ssec:floquet}

In this subsection we provide a detailed derivation of the Floquet eigensystem, consisting of the Floquet exponents and left/right eigenvectors, which are employed in the analysis of limit cycle diffusion. We consider the system of $N$ linear first order ODEs given by Eq.~(E5) of the manuscript:
\begin{align}
    \dot{\vec{z}} = \mathbf{J}(t)\vec{z}
    \label{eq:floquetSys}
\end{align}
A key constraint of the problem is that the dynamical matrix $\mathbf{J}(t)$ is periodic: $\mathbf{J}(t) = \mathbf{J}(t+T)$. Being a system of $N$ ODEs, it admits $N$ linearly independent solutions which we label $\{\vec{z}_1(t),\vec{z}_2(t),\ldots,\vec{z}_N(t)\}$. We can construct a matrix $\mathbf{R}(t)$ with linearly independent columns $\{\vec{z}_i(t)\}$; the resulting matrix also satisfies:
\begin{align}
    \dot{\mathbf{R}}(t) = \mathbf{J}(t)\mathbf{R}(t)
    \label{eq:floquetR}
\end{align}
Being a linear system, multiplying $\mathbf{R}(t)$ by a constant matrix $\mathbf{K}$ also satisfies the system. In particular, if we define the matrix $\mathbf{V}(t)$ as:
\begin{align}
    \mathbf{V}(t) = \mathbf{R}(t)\mathbf{K}
\end{align}
then:
\begin{align}
    \dot{\mathbf{V}}(t) = \dot{\mathbf{R}}(t)\mathbf{K} = \mathbf{J}(t)\mathbf{R}(t)\mathbf{K} = \mathbf{J}(t)\mathbf{V}(t) 
\end{align}
so that $\mathbf{V}(t)$ is also a solution of the Floquet system. Since $\mathbf{J}(t+T) = \mathbf{J}(t)$, we find:
\begin{align}
    \dot{\mathbf{R}}(t+T) = \mathbf{J}(t)\mathbf{R}(t+T)
\end{align}
so that the matrix $\mathbf{R}(t+T)$ also solves the linear system. Combining the above two results, we can relate $\mathbf{R}(t+T)$ to $\mathbf{R}(t)$:
\begin{align}
    \mathbf{R}\left( t+T \right) = \mathbf{R}(t)\mathbf{K}
\end{align}
Since $\mathbf{K}$ is a constant matrix, it can be obtained from the above relation by setting $t=0$:
\begin{align}
    \mathbf{K} = \mathbf{R}^{-1}(0)\mathbf{R}(T)
\end{align}
By choosing initial conditions such that $\mathbf{R}(0) = \mathbf{I}$, then we simply obtain $\mathbf{K} = \mathbf{R}(T)$, which is a constant matrix referred to as the \textit{fundamental matrix} of the Floquet system. It is obtained by solving Eq.~(\ref{eq:floquetR}) for $\mathbf{R}(t)$ as a function of time $t \in [0, T]$ over a single period $T$ of the classical solution, with the aforementioned initial condition.

Note that the fundamental matrix is in general non-Hermitian; as such we need to consider its complex eigenvalues $\rho_i$ and left/right eigenvectors $\vec{b}_i$, $\vec{c}_i$ respectively:
\begin{align}
    \mathbf{K} \vec{b}_i &= \rho_i \vec{b}_i \nonumber \\
    \vec{c}_i^{\dagger} \mathbf{K} &= \rho_i \vec{c}_i^{\dagger}
\end{align}
which satisfy the orthogonality relation:
\begin{align}
    \vec{c}_i^{\dagger}\vec{b}_j = \delta_{ij}
\end{align}
If we now define the set of vectors $\{\vec{y}_i(t)\}$:
\begin{align}
    \vec{y}_i(t) = \mathbf{R}(t)\vec{b}_i
\end{align}
Substituting the above into Eq.~(\ref{eq:floquetSys}), we find:
\begin{align}
    \dot{\vec{y}}_i(t) = \dot{\mathbf{R}}(t)\vec{b}_i = \mathbf{J}(t)\mathbf{R}(t)\vec{b}_i = \mathbf{J}\vec{y}_i(t)
\end{align}
which means $\{\vec{y}_i(t)\}$ are solutions to the Floquet system, Eq.~(\ref{eq:floquetSys}), as well. This decomposition of the solutions in terms of the eigenvectors of the fundamental matrix also implies:
\begin{align}
    \vec{y}_i(t+T) &= \mathbf{R}(t+T)\vec{b}_i = \mathbf{R}(t)\mathbf{K}\vec{b}_i = \rho_i \mathbf{R}(t)\vec{b}_i \nonumber \\
    &= \rho_i \vec{y}_i(t) 
\end{align}
Therefore solutions to the Floquet system are in general \textit{not} periodic, unless $\rho_i = 0$. The set of $\{\rho_i\}$ are referred to as \textit{Floquet multipliers}. However, the solutions separated by a period are simply related by a constant. In particular, this enables writing them in the form:
\begin{align}
    \vec{y}_i(t) = e^{\mu_i t}\vec{p}_i(t)
\end{align}
where we introduce a set of periodic vectors $\{\vec{p}_i(t)\}$, such that:
\begin{align}
    \vec{p}_i(t+T) = \vec{p}_i(t)
\end{align}
Then, 
\begin{align}
    \vec{y}_i(t+T) = e^{\mu_i T} e^{\mu_i t} \vec{p}_i(t+T) = e^{\mu_i T} e^{\mu_i t} \vec{p}_i(t) = e^{\mu_i T} \vec{y}_i (t) \equiv \rho_i \vec{y}_i(t)
\end{align}
This enables a parameterization of the Floquet multipliers in terms of \textit{Floquet exponents} $\{\mu_i\}$:
\begin{align}
    \rho_i = e^{\mu_i T}
\end{align}
We define the periodic vectors $\{\vec{p}_i(t)\}$ as the Floquet right eigenvectors. They are completely determined by the eigenvectors $\{\vec{b}_i\}$ and eigenvalues $\{\rho_i\}$ of the fundamental matrix via:
\begin{align}
    \vec{y}_i(t) = \mathbf{R}(t)\vec{b}_i = e^{\mu_i t} \vec{p}_i(t) \implies \vec{p}_i(t) = e^{-\mu_i t}\mathbf{R}(t)\vec{b}_i
    \label{eq:pEq}
\end{align}
To find the equation of motion for the Floquet right eigenvectors $\vec{p}_i(t)$, we can simply take the time derivative of the above, which then yields the equation of motion:
\begin{align}
    \dot{\vec{p}}_i(t) &= \left[ -\mu_i  + \mathbf{J}(t) \right] e^{-\mu_i t}\mathbf{R}(t) \vec{b}_i \nonumber \\
    \implies \dot{\vec{p}}_i(t) &= \left[ \mathbf{J}(t) - \mu_i \right] \vec{p}_i(t)
\end{align}

Similar to the definition of $\vec{y}_i(t)$, we can define solutions $\{\vec{w}_i(t)\}$ in terms of the left eigenvectors of the fundamental matrix $\mathbf{R}(t)$:
\begin{align}
    \vec{w}_i^{\dagger}(t) = \vec{c}_i^{\dagger}\mathbf{R}^{-1}(t)
\end{align}
Clearly, we have:
\begin{align}
    \vec{w}_i(t+T) = \vec{c}_i^{\dagger} \mathbf{R}^{-1}(t+T) = \vec{c}_i^{\dagger} \mathbf{K}^{-1}\mathbf{R}^{-1}(t) = \rho_i^{-1}\vec{c}_i^{\dagger}\mathbf{R}^{-1}(t) = \rho_i^{-1}\vec{w}_i(t)
\end{align}
where we have used the relationship:
\begin{align}
    \vec{c}_i^{\dagger}\mathbf{K} = \rho_i \vec{c}_i^{\dagger} \implies \vec{c}_i^{\dagger} = \rho_i \vec{c}_i^{\dagger}\mathbf{K}^{-1} \implies \vec{c}_i^{\dagger}\mathbf{K}^{-1} = \rho_i^{-1}\vec{c}_i^{\dagger}
\end{align}
Then, since $\rho_i = e^{\mu_i t}$, we can write $\vec{w}_i(t)$ in terms of a periodic vector $\vec{q}_i(t) = \vec{q}_i(t+T)$:
\begin{align}
    \vec{w}_i(t) = e^{-\mu_i t}\vec{q}_i^{\dagger}(t)
\end{align}
We analogously define the set of periodic vectors $\{\vec{q}_i(t)\}$ as the left Floquet eigenvectors, which are again completely determined by the eigenvalues and eigenvectors of the fundamental matrix as:
\begin{align}
    \vec{q}_i^{\dagger}(t) = e^{\mu_i t}\vec{c}_i^{\dagger}\mathbf{R}^{-1}(t)
    \label{eq:qEq}
\end{align}

We can also determine an equation of motion for the Floquet left eigenvectors by taking the time derivative of the above relation. This requires the time derivative of the inverse of $\mathbf{R}(t)$:
\begin{align}
    \frac{d}{dt}(\mathbf{R}\mathbf{R}^{-1}) &= \frac{d}{dt}\mathbf{I} = 0 = \dot{\mathbf{R}}\mathbf{R}^{-1} + \mathbf{R}\dot{\mathbf{R}}^{-1} \nonumber \\
    \implies \dot{\mathbf{R}}^{-1} &= - \mathbf{R}^{-1}\dot{\mathbf{R}}\mathbf{R}^{-1} \nonumber \\
    \implies \dot{\mathbf{R}}^{-1} &= -\mathbf{R}^{-1}\mathbf{J}
\end{align}
which then yields the equation of motion:
\begin{align}
    \dot{\vec{q}}_i^{\dagger}(t) &= e^{\mu_i t}\vec{c}_i^{\dagger}\mathbf{R}^{-1}(t)\left[ \mu_i - \mathbf{J}  \right] \nonumber \\
    \implies \dot{\vec{q}}_i^{\dagger}(t) &= \vec{q}_i^{\dagger}(t)\left[ \mu_i - \mathbf{J}(t)  \right]
\end{align}

Finally, we note that the right and left Floquet eigenvectors satisfy the orthogonality relationship:
\begin{align}
    \vec{q}_j^{\dagger}(t)\vec{p}_i(t) = e^{(\mu_j-\mu_i)t}\vec{c}_j^{\dagger}\vec{b}_i = \delta_{ij}
    \label{eq:floquetNorm}
\end{align}
at all times $t \in [0, T]$, as can be easily found from the definitions of the Floquet eigenvectors, Eqs.~(\ref{eq:pEq}),~(\ref{eq:qEq}).

Finally, we show here that provided the Floquet system admits a periodic solution, at least one of the Floquet exponents vanishes~\cite{haken_at_1983}. The corresponding Floquet eigenvector is then proportional to the velocity vector of the limit cycle solution. To do so, we begin with Eq.~(E2) from the manuscript that describes the limit cycle velocity:
\begin{align}
    \vec{v} = \frac{d\vec{\zeta}_{\rm c}}{dt} = \vec{A}_{\rm c}[\vec{\zeta}_c(t)]
\end{align}
Differentiating the above with respect to time, we obtain:
\begin{align}
    \dot{\vec{v}}^i = \sum_{j} \partial_j \vec{A}^i_{\rm c}[\vec{\zeta}_{\rm c}(t)] \vec{v}^j \implies \dot{\vec{v}} = \mathbf{J}[\vec{\zeta}_c(t)]\cdot \vec{v} \implies \dot{\vec{v}} = \left[\mathbf{J}(t) - 0\right]\vec{v}
\end{align}
where we have used the chain rule since $\vec{A}_{\rm c}[\vec{\zeta}_c(t)]$ depends on time only via its dependence on $\vec{\zeta}_{\rm c}(t)$, and we recall that the Jacobian matrix elements are defined via $J_{ij} = \partial_j A_{\rm c}^i$. When written as the third term, it becomes clear that $\vec{v}$ satisfies the equation of motion for the Floquet eigenvector $\vec{p}_i(t)$ with $\mu_i = 0$, Eq.~(\ref{eq:pEq}). We label this Floquet exponent with index 0, $\mu_0 = 0$. Clearly, the corresponding eigenvector $\vec{p}_0$ is then proportional to the tangential velocity $\vec{v}$, differing only by a constant that is set by the normalization requirement for the left and right eigenvectors, Eqs.~(\ref{eq:floquetNorm}).

\begin{center}
    \rule{10cm}{1pt}
\end{center}


\end{document}